\documentclass[11pt, a4paper]{article}

\usepackage[utf8]{inputenc}
\usepackage{jheppub}
\usepackage{tikz}
\usetikzlibrary{positioning}
\tikzset{font=\small, >=stealth, shorten >=0.1cm, shorten <=0.1cm}

\newcommand*\circled[1]{\tikz[baseline=(char.base)]{
            \node[shape=circle,draw,inner sep=2pt] (char) {#1};}}

\newcommand{\Tr}{\mathop{\mathrm{Tr}}\nolimits}
\newcommand{\mD}{\mathcal{D}}
\newcommand{\mA}{\mathcal{A}}

\usepackage{dsfont}
\usepackage{enumitem}
\usepackage{mathtools}

\usepackage{slashed}

\allowdisplaybreaks[1]

\title{Four-Dimensional $\mathcal{N}=2$ Supersymmetric Theory with Boundary as a Two-Dimensional Complex Toda Theory}

\author{Yuan Luo,}
\author{Meng-Chwan Tan,}
\author{Petr Vasko,}
\author{and Qin Zhao}

\emailAdd{tongxueluo@gmail.com, mctan@nus.edu.sg, \\ petr.vasko@fuw.edu.pl,
zhaoqin@nus.edu.sg}

\affiliation{Department of Physics, National University of
  Singapore \\
2 Science Drive 3, Singapore 117551}

\abstract{We perform a series of dimensional reductions of the 6d, $\mathcal{N}=(2,0)$ SCFT on    $S^2\times\Sigma\times I\times S^1$ down to 2d on $\Sigma$. The reductions are performed in three steps: (i) a reduction on $S^1$ (accompanied by a topological twist along $\Sigma$) leading to a supersymmetric Yang-Mills theory on $S^2\times\Sigma\times I$, (ii) a further reduction on $S^2$ resulting in a complex Chern--Simons theory defined on $\Sigma\times I$, with the real part of the complex Chern-Simons level being zero, and the imaginary part being proportional to the ratio of the radii of $S^2$ and $S^1$, and (iii) a final reduction to the boundary modes of complex Chern--Simons theory with the Nahm pole boundary condition at both ends of the interval $I$, which gives rise to a complex Toda CFT on the Riemann surface $\Sigma$. As the reduction of the 6d theory on $\Sigma$ would give rise to an $\mathcal{N}=2$ supersymmetric theory on $S^2\times I\times S^1$, our results imply a 4d-2d duality between four-dimensional $\mathcal{N}=2$ supersymmetric theory with boundary and two-dimensional complex Toda theory.}


\arxivnumber{}

\begin{document}
\maketitle

\flushbottom

\section{Introduction, Summary of Results and Relevant Works}

The two main classes of dualities that can be derived from dimensional reductions of the 6d, $\mathcal{N}=(2,0)$ theory come under the names of 4d-2d \cite{Tachikawa:2016kfc} and 3d-3d \cite{Dimofte:2016pua} dualities. The state of the art in this area of research is nicely summarized in \cite[Introduction]{Assel:2016lad}, which also comes with a comprehensive list of references. In this paper, we would like to derive a 4d-2d duality that has not yet been listed in \cite[Introduction]{Assel:2016lad}. 

To this end, note that in \cite{Bawane:2014uka}, a supersymmetric theory was constructed on $S^2\times S^2$, and its partition function was computed using the technique of localization. The authors observed that the building blocks of the partition function contained three-point functions and conformal blocks of Liouville gravity \cite{Zamolodchikov:2005fy}, \cite{Belavin:2005jy}, and so they conjectured a novel AGT-like correspondence between the 4d supersymmetric theory on $S^2\times S^2$ and 2d Liouville gravity. 

To obtain a proof of this conjecture, one will need to consider two physically equivalent reductions of the $\mathcal{N}=(2,0)$ SCFT on $M_6=S^2\times\Sigma\times S^2$. The first reduction on $\Sigma$ presumably gives us a 4d supersymmetric theory on $S^2\times S^2$, while the second reduction on $S^2\times S^2$ ought to give us Liouville gravity on $\Sigma$. Our original goal was to explicitly carry out the second reduction along the lines of \cite{Cordova:2016cmu}. 

However, it seems that the geometry of $M_6=S^2\times\Sigma\times S^2$ does not satisfy the constraints imposed by integrability conditions of the generalized Killing spinor equations, i.e. there are no preserved supersymmetries on $M_6$. (What we can say is that a rather general ansatz motivated by group theory considerations did not pass the test, although we have to emphasize that we do not have a no-go theorem, since it is quite involved to handle a full ansatz for the self-dual three form $T$ from the six dimensional viewpoint, especially due to the appearance of non-linear terms.) 

That being said, note that the metric of $M_6=S^2\times\Sigma\times S^2$ can be written as
\begin{equation}
ds^2=R^2\left(d\theta^2+sin(\theta)^2d\phi^2\right)+e^{A(x,y)}\left(dx^2+dy^2\right)+\widetilde{R}^2\left(d\widetilde{\theta}^2+sin(\widetilde{\theta})^2d\widetilde{\phi}^2\right),
\end{equation}
where $A(x,y)$ is a local conformal factor of the metric on the Riemann surface $\Sigma$ with coordinates $x,y$; in the cylinder limit $sin(\widetilde{\theta})\to\mathrm{const}$, we have $M_6=S^2\times\Sigma\times I\times S^1$, and in this case, we {\it have} preserved supersymmetries. As such, we shall, in this paper, consider the reduction of the $\mathcal{N}=(2,0)$ SCFT in this cylinder limit.

\bigskip\noindent{\it Summary of Results}

This paper is devoted to dimensionally reducing the 6d, $\mathcal{N}=(2,0)$ superconformal field theory (SCFT) on the manifold $M_6=S^2\times\Sigma\times I\times S^1$ (where we topologically twist along $\Sigma$, a Riemann surface) over $S^2\times I\times S^1$ down to $\Sigma$. The reduction is performed in {three steps}, and the {main results} of this work can be summarized as follows:
\begin{enumerate}[label=(\roman*),ref=(\roman*)]
 \item We first reduce on $S^1$ to obtain a five-dimensional supersymmetric Yang-Mills (SYM) theory defined on $M_5=S^2\times\Sigma\times I$, which retains all the original sixteen supersymmetries. It is specified by the action provided in \eqref{eq:SAsub}--\eqref{eq:Sintsub}. (The corresponding sixteen supercharges explicitly form a superconformal algebra, and can be regarded as a conformal extension of eight supercharges. This point will be further elaborated at the end of section 3.1.) \label{it:SYM}
 \item Next, we dimensionally reduce the obtained 5d SYM on $S^2$ to get a complex Chern--Simons theory on $M_3=\Sigma\times I$, with the quantized real part of the level $k=0$ and the imaginary part of the level $s=i\tfrac{R}{\widetilde{R}}$ (where $R$ [$\widetilde{R}$] is the radius of $S^2$ [$S^1$]). The action is given in \eqref{csaction}.\label{it:cCS} 

 \item Lastly, we study the boundary modes of the complex Chern--Simons theory. They turn out to describe a complex Toda CFT on $\Sigma$ with coupling constant $\left(k=0, s=i\tfrac{R}{\widetilde{R}}\right)$. The action is given in \eqref{complex_liouville}. \label{it:Toda}

\end{enumerate}

  Then, by our results listed above, we will conclude the paper by showing that there exists a 4d-2d duality between four-dimensional $\mathcal{N}=2$ supersymmetric theory with boundary on $S^2 \times I \times S^1$ and two-dimensional complex Toda theory on $\Sigma$.

\bigskip\noindent{\it Relevant Works}

Since the 6d $\mathcal{N}=(2,0)$ SCFT has no known Lagrangian description, step \ref{it:SYM} is essential in the sense that only after which, we have a Lagrangian theory to which we can apply standard field theory methods. The appropriate framework for this dimensional reduction was developed in \cite{Cordova:2013bea}. It is applicable whenever the six dimensional manifold $M_6$ contains a (possibly non-trivial) circle fibration, $S^1\to M_6\to M_5$. In such a case, the authors of \cite{Cordova:2013bea} showed that the resulting theory defined on $M_5$ is a supersymmetric Yang-Mills theory in general supergravity background; they provided explicit expressions determining this theory, in particular, its Lagrangian. It depends on bosonic supergravity fields, and these are either fixed by geometry of the fibration $S^1\to M_6\to M_5$ or have to be derived by solving the generalized Killing spinor equations. These equations are indeed the central object of the whole reduction procedure, since (I) they allow to solve for bosonic supergravity fields that are not already fixed by geometry, and (II) they provide generalized Killing spinors parameterizing supersymmetry transformations of the SYM theory, i.e. supersymmetry is preserved only if they admit non-trivial solutions or equivalently when integrability conditions are fulfilled, which in turn imposes constraints on the geometry of $M_5$.

We should state that conclusions similar to that obtained in step \ref{it:cCS} were derived earlier in \cite{Yagi:2013fda} and \cite{Lee:2013ida} using different methods.\footnote{In these references, a complex Chern--Simons theory with level $k=0\ \text{and}\ s=8\pi^2\tfrac{R}{e^2}$, where $e^2\propto\widetilde{R}$, was obtained. The geometric dependence of the imaginary part of the level $s$ on the radii of the sphere and the circle is consistent with our result $s=i\tfrac{R}{\widetilde{R}}$. However, note that our level $s$ is purely imaginary while theirs is real. Nevertheless, 
both cases correspond to a unitary branch of complex Chern--Simons theory as discussed in \cite{Witten:1989ip}.} The authors used a bottom-up approach (based partly on trial and error methods) to construct 5d SYM on $M_5$, and then applied supersymmetric localization to get the Chern-Simons theory with level $k=0$ on $M_3$. Nonetheless, it is still useful to rederive their results from our top-down approach --  this way, we can exploit the power of the formalism introduced in \cite{Cordova:2013bea} and test its validity.

In fact, the formalism in \cite{Cordova:2013bea}  has been already applied in \cite{Cordova:2013cea}, \cite{Cordova:2016cmu} and \cite{Assel:2016lad}. These works study reductions of the 6d $\mathcal{N}=(2,0)$ SCFT formulated on different six-manifolds $M_6$, and all are related to topics investigated in this paper. For instance, \cite{Assel:2016lad} explains how to properly take care of (Nahm pole) boundary conditions -- this is precisely what we need, since our geometry contains a boundary almost identical to theirs. In \cite{Cordova:2013cea} and \cite{Cordova:2016cmu}, the authors considered $\mathcal{N}=(2,0)$ SCFT on $S^3/\mathbb{Z}_k\times M_3$, where $k\in\mathbb{Z}_{>0}$, and reduced it to complex Chern--Simons theory on $M_3$ with quantized level $k$ -- thus, our result in step \ref{it:cCS} is parallel to theirs where we obtain complex Chern--Simons theory at level $k=0$ instead because we consider $S^2  \times S^1$ and not $S^3/\mathbb{Z}_k$. 

The appropriate objects suited to capture the general situation are lens spaces $L(k,1)$. Consider the $\mathcal{N}=(2,0)$ SCFT on $M_6=L(k,1)\times M_3$. Dimensional reduction on $L(k,1)$ results in complex Chern--Simons theory on $M_3$ at level $(k,s)$. The quantized real part $k$ is fixed by the geometric invariant specifying the lens space, while the imaginary part $s$ is also fixed by geometric considerations, although it depends on the details of the fibration $S^1\to M_6\to M_5$ whence it cannot be described uniformly. Setting $k=0$ reduces to the geometry we are dealing with in this paper, since $L(0,1)\simeq S^2\times S^1$, and one gets $s=i\dfrac{R}{\widetilde{R}}$. For $k\geq 1$, the lens space $L(k,1)$ is isomorphic to $S^3/\mathbb{Z}_k$, and $s=0$ for the non-squashed space, while for the squashed one, $S^3_\ell/\mathbb{Z}_k$, $s=\sqrt{1-\ell^2}$.

Although \cite{Cordova:2016cmu} provided the generalization to $k>1$, the main corollary of that paper was to explain how Toda CFT emerges in the AGT correspondence \cite{Alday:2009aq}. This was done by considering dimensional reduction on $S^4$ (regarded as a nontrivial product of $S^3$ and $I$) of the $\mathcal{N}=(2,0)$ SCFT defined on $S^4\times\Sigma$, and explicitly showing that it leads to real Toda theory on $\Sigma$. The 
essence of this analysis was a careful treatment of boundary modes of complex Chern--Simons theory and their reduction to Toda fields caused by Nahm pole boundary conditions. In this sense, the conclusions obtained in step \ref{it:Toda} are just a specialization of the more general results gathered in \cite{Cordova:2016cmu}. As the reduction of  complex Chern--Simons theory on $M_3 = \Sigma \times I$ to complex Toda theory on $\Sigma$ holds for all $k$, all that is needed in our case is to consider a $k=0$ complex Chern--Simons theory in place of a $k\geq1$ one. It is only when one tries to relate the complex Toda theory to a real Toda theory plus (possibly) a decoupled CFT,  as was done in \cite{Cordova:2016cmu}, that the $k=0$ case, unlike the $k \geq 1$ case, encounters an obstruction that we will elaborate upon in Section \ref{S:Toda}.

\bigskip\noindent{\it Plan of the Paper}

The plan of the paper is as follows. In Section \ref{S:geometry} we consider the 6d $\mathcal{N}=(2,0)$ SCFT on $M_6=S^2\times\Sigma\times I\times S^1$, fully describe the geometry of the base $M_5=S^2\times\Sigma\times I$ of the circle fibration, and provide an ansatz for bosonic supergravity background fields based on group theory methods. In Section \ref{S:GKSE},  we use this ansatz to simplify and solve the generalized Killing spinor equations. Finishing this step completely determines the 5d SYM theory. In Section \ref{S:reduction}, we move on to examine the dimensional reduction of the 5d SYM on $S^2$. This analysis culminates by recognizing that the three dimensional effective theory is a complex Chern--Simons theory. Finally, in Section \ref{S:Toda}, we briefly discuss that the theory living on the boundaries of $\Sigma\times I$ with Nahm pole boundary conditions for the complexified Chern--Simons connection at both ends of the interval, can be obtained along the lines of \cite{Cordova:2016cmu} in an unaltered way. The result is a complex Toda theory defined on $\Sigma$. 

Index conventions, gamma matrices and technical expressions displaying results of dimensional reductions are summarized in three short appendices.

\section{Supergravity background fields on $S^2\times \Sigma \times I $}\label{S:geometry}
In this section, we introduce the space-time structure of six-dimensional (2,0) supergravity and discuss the transformation properties of the supergravity background fields under the Lorentz symmetry and R-symmetry. 

To begin with, the 6d metric on $S^2\times \Sigma \times I \times S^1$ is given by
\begin{equation}\label{eq:line_el}
ds^2 = R^2(d\theta^2 +\sin^2\theta d\phi^2)+e^{A(x,y)}\left(dx^2 + dy^2\right) +du^2+\tilde{R}^2d\tilde{\phi}^2,
\end{equation}
where $A(x,y)$ is a conformal factor of the metric on $\Sigma$ while $R$ and $\tilde{R}$ are the radii of the $S^2$ and the $S^1$, respectively. In a general six-dimensional (2,0) supersymmetric theory, the Lorentz and R-symmetry group is $SO(6)_L\times Sp(4)_R$. To describe our background, it is natural to split the Lorentz group as $SO(2)_{S^2} \times SO(2)_{\Sigma}\times SO(2)_{S^1\times I}$. However, since there is no Lagrangian for six-dimensional SCFT, our strategy is to first construct the five-dimensional SYM on $S^2\times \Sigma \times I$ in a supergravity background by dimensional reduction of the six-dimensional theory on the $S^1$. Hence, the remaining Lorentz group is $SO(2)_{S^2} \times SO(2)_{\Sigma}$.\footnote{One may note that in the bulk of the three-dimensional manifold $\Sigma\times I$, Lorentz symmetry $SO(3)$ can be restored, and only the  boundary of the interval $I$ breaks the Lorentz symmetry, which results in the residual Lorentz symmetry $SO(2)$ on $\Sigma$.
 } Correspondingly, we split the R-symmetry group $Sp(4)_R$ as $SO(2)_R\times SO(2)_R$. 

Given the decompositions of the symmetry groups, we now provide the transformation properties of the background fields under these decomposed groups, which can guide us to choose the correct ansatz for the background fields and simplify the Killing spinor equations further in the next section. First, a summary of the background fields for the six-dimensional off-shell gravity multiplet is shown in Table \ref{boson_back}, where the conventions for indices are summarized in Appendix \ref{conventions}.

From the line element $ds^2=\delta_{\underline{A}\underline{B}}e_{\underline{\mu}}^{\underline{A}}e_{\underline{\nu}}^{\underline{B}}dx^{\underline{\mu}}dx^{\underline{\nu}}$ given in \eqref{eq:line_el} we extract the coframe $e_{\underline{\mu}}^{\underline{A}}$. Since we will be using the framework of \cite{Cordova:2013bea}, we need to compare to their more general coframe $E_{\underline{\mu}}^{\underline{A}}$ (adapted to a circle fibration over a five dimensional manifold), in order to identify some of the supergravity background fields already at this stage. The coframes read
\begin{equation}
e_{\underline{\mu}}^{\underline{A}}=\mathrm{diag}\left(R,R\sin\theta,e^{\frac{1}{2}A(x,y)},e^{\frac{1}{2}A(x,y)},1,\tilde{R}\right),\quad E^{\underline{A}}_{\underline{\mu}} = \left( 
\begin{array}{ll}
E^A_\mu & E^6_\mu = \alpha^{-1} C_\mu= 0 \\ E_6^A=0 & E_6^6 =\alpha^{-1}\end{array} \right),
\end{equation} 
from where we conclude that the supergravity background fields -- the scalar field $\alpha$ and the five-dimensional gauge field $C_\mu$ (called the dilaton and graviphoton, respectively) -- take the values
\begin{equation}
\alpha=\tilde{R}^{-1},\quad C_\mu=0.
\end{equation}
Therefore, the dilaton is a constant field and the field strength $G=dC$ for the graviphoton vanishes, which simplifies the generalized Killing spinor equations studied later. The remaining bosons are reduced as 
\begin{equation}
\begin{array}{lll}
V_{\underline{A}\hat{B}\hat{C}} &\rightarrow & \left\lbrace \begin{array}{l}
V_{A\hat{B}\hat{C}} \\
V_{6\hat{B}\hat{C}} \equiv S_{\hat{B}\hat{C}}
\end{array} \right., \\
T_{\hat{A}\underline{BCD}} & \rightarrow & T_{\hat{A}BC6} \equiv T_{\hat{A}BC},\\
D_{(\hat{A}\hat{B})} & \rightarrow & D_{(\hat{A}\hat{B})}.
\end{array}
\end{equation}

\begin{table}
\caption{The bosonic gauge fields of the 6d $(2,0)$ conformal supergravity}\label{boson_back}
\begin{center}
\begin{tabular}{|c|c|c|c|c|}
\hline
Label & Field & Properties & $so(6)_L$  & $sp(4)_R$\\
\hline
$e^{\underline{A}}_{\underline{\mu}}$ & Coframe  & -- & 6 & 1 \\
$\alpha$ & Dilaton & --  & 1 & 1 \\
$V_{\underline{A}\hat{B}\hat{C}}$ & R-symmetry gauge field & $V_{\underline{A}\hat{B}\hat{C}}=-V_{\underline{A}\hat{C}\hat{B}}$ & 6 & 10 \\
$T_{\hat{A}[\underline{BCD}]}$ & Auxiliary 3-form & $T_{\hat{A}}= - \star T_{\hat{A}}$ & 10 & 5 \\
$D_{(\hat{A}\hat{B})}$ & Auxiliary scalar & $D_{\hat{A}\hat{B}}= D_{\hat{B}\hat{A}}$, $D^{\hat{A}}_{\hat{A}}=0$ & 1 & 14 \\
\hline
\end{tabular}
\end{center}
\end{table}

Under the Lorentz symmetry group before and after the splitting we performed, the background fields transform as
\begin{equation}
\begin{array}{rrrl}
& SO(6)_L & \rightarrow & SO(2)_{S^2}\times SO(2)_{\Sigma} \\
\underline{A}:& 6 & \rightarrow & (\pm 2, 0)\oplus(0, \pm 2) \oplus 2\times (0,0)\\ 
\quad [\underline{BCD}]^{(+)}: & 10 &\rightarrow & 2 \times (0,0) \oplus (\pm 2, \pm 2) \oplus (\pm 2, 0) \oplus (0,\pm 2) \\
\quad [\underline{BC}]: & 15 & \rightarrow & 3\times (0,0) \oplus (\pm 2, \pm 2) \oplus 2\times (\pm 2, 0) \oplus 2\times (0,\pm 2)
\end{array}
\end{equation}
and under the R-symmetry group as 
\begin{equation}\label{r-symmetry}
\begin{array}{rrrl}
& SO(5)_R& \rightarrow & SO(2)_{R}\times SO(2)_{R} \\
\hat{A}:& 5 & \rightarrow & (\pm 2, 0)\oplus(0, \pm 2) \oplus  (0,0) \\
\quad [{\hat{B}\hat{C}}]: & 10 &\rightarrow & 2 \times (0,0) \oplus (\pm 2, \pm 2) \oplus (\pm 2, 0) \oplus (0,\pm 2) \\
(\hat{A}\hat{B}): & 14 & \rightarrow & (\pm 4, 0) \oplus (0, \pm 4) \oplus (\pm 2, \pm 2) \oplus (\pm 2, 0)\oplus (0,\pm 2) \oplus 2\times( 0,0).
\end{array}
\end{equation}

Now, we twist our theory on $\Sigma$ by defining the new Lorentz group $SO(2)'_\Sigma$ as the diagonal subgroup of $SO(2)_\Sigma \times SO(2)_R$. Then, under the residual symmetry group $SO(2)_{S^2}\times SO(2)_R \times SO(2)'$, the background fields $T_{\hat{A}[\underline{BCD}]}$, $V_{\underline{A}\hat{B}\hat{C}}$, and $D_{\hat{A}\hat{B}}$ transform respectively as 
\begin{equation}
\begin{array}{rll}
SO(6)_L \times SO(5)_R & \rightarrow & SO(2)_{S^2}\times SO(2)_R \times SO(2)' \\
T_{\hat{A}\underline{BCD}}:\quad (10, 5) & \rightarrow & (\pm 2, \pm 2, \pm 2) \oplus (\pm 2, 0, \pm 4) \oplus 2\times(\pm 2, 0, \pm 2) \oplus (\pm 2, \pm 2, 0) \oplus (0, \pm 2, \pm 2) \\
& & \oplus (0, 0, \pm 4) \oplus 3\times(\pm 2, 0, 0) \oplus 3\times(0, 0, \pm 2) \oplus 2\times (0, \pm 2, 0) \oplus  4\times(0,0, 0)\\
V_{\underline{A}\hat{B}\hat{C}}:\quad (6, 10) & \rightarrow & (\pm 2, \pm 2, \pm 2)\oplus (0,\pm 2,\pm 4)\oplus 3\times (0,\pm 2,\pm 2) \oplus (\pm 2,\pm 2,0) \\
& & \oplus (\pm 2,0, \pm 2)\oplus (0, 0, \pm 4) \oplus 4\times (0,\pm 2, 0) \oplus 4\times (0, 0, \pm 2) \\
& & \oplus 2\times (\pm 2,0,0) \oplus 6\times (0,0,0) \\
D_{(\hat{A}\hat{B})}:\quad (1, 14) & \rightarrow & (0, \pm 2, \pm 2) \oplus (0, \pm 4, 0)\oplus (0,0,\pm 4) \oplus (0, 0,\pm 2) \oplus (0, \pm 2, 0) \oplus  2\times(0, 0, 0).
\end{array}
\end{equation}
Based on the fact that the background fields should be singlets in 5d theory, we summarize the possible non-vanishing background fields as follows:

\begin{align}\label{eq:SUGRAansatz}
\begin{aligned}[c]
S_{\hat{1}\hat{2}}&=s \\
S_{\hat{3}\hat{4}}&=\widetilde{s}
\end{aligned}
&&
\begin{aligned}[c]
V_{3\hat{4}\hat{5}}&=w \\
V_{4\hat{3}\hat{5}}&=\widetilde{w} \\
V_{5\hat{1}\hat{2}}&=v \\
V_{5\hat{3}\hat{4}}&=\widetilde{v}
\end{aligned}
&&
\begin{aligned}[c]
T_{\hat{3}45}&=\widetilde{T} \\
T_{\hat{4}35}&=T \\
T_{\hat{5}12}&=t \\
T_{\hat{5}34}&=\widetilde{t},
\end{aligned}
\end{align}
while the general ansatz for $D$ is $D_{\hat{A}\hat{B}}= d_1(\delta_{\hat{1}\hat{1}}+\delta_{\hat{2}\hat{2}})+d_2(\delta_{\hat{3}\hat{3}}+\delta_{\hat{4}\hat{4}})+2(-d_1-d_2)\delta_{\hat{5}\hat{5}}$.
%
%

\section{Solving the generalized Killing spinor equations}\label{S:GKSE}

The generalized Killing spinor equations (GKSE) of six dimensional $N=(2,0)$
supergravity reduced to five dimensions were obtained in \cite{Cordova:2013bea}.
We briefly review the logic behind their derivation. In order to construct a
quantum field theory defined on a rather general~\footnote{Constraints on the
  geometry of the manifold come from integrability conditions for the system of
  generalized Killing spinor equations.} five dimensional manifold $M_5$ that is
invariant under rigid supersymmetry, we use supergravity as a tool to achieve it
\cite{Festuccia:2011ws}. Concretely, the $(2,0)$ abelian tensor multiplet in six dimensions was coupled to $(2,0)$ conformal supergravity \cite{Bergshoeff:1999db,Riccioni:1997np} in six dimensions. Finally, this coupled system was dimensionally reduced to five dimensions and generalized to a non-abelian setting, obtaining thus a non-abelian supersymmetric Yang-Mills theory defined on $M_5$ that admits solutions to GKSE (otherwise supersymmetry of the resulting theory would be broken) \cite{Cordova:2013bea}.

The above construction especially implies that bosonic supergravity fields are
non-dynamical, i.e. should be treated as background fields. In general, they are functions on $M_5$. However, it happens quite often that they actually take constant values related to geometrical invariants of the five dimensional
(pseudo)-Riemannian manifold (for example they can be proportional to the scalar curvature). As we already emphasized the bosonic supergravity fields are fixed and therefore, their variation under local
supersymmetry transformations (proportional to fermions in the supergravity
theory) has to vanish, implying that all fermion fields have to take zero
values. Thus, applying a local supersymmetry transformation $\delta$ to all the
fermion fields has to give a vanishing result to keep them at zero value
\begin{equation}
  \delta\left(\mathrm{fermions}\right)=0. 
\end{equation}
These conditions are the generalized Killing spinor equations for
spinors parameterizing rigid supersymmetry variations in the 5d SYM theory. They
take the form (note that the spinor index of $Spin(5)$ is suppressed)
\begin{align}\label{eq:GKSE1}
  0=\delta \psi_{A}^{m}=&\mathcal{D}_{A}\zeta^{m}+\frac{i}{2\alpha}\left[\phantom{\frac{1}{1}}\hspace{-.11in}G_{AB}\Omega^{mn}-\alpha S^{mn}\eta_{AB}\right]\Gamma^{B}\zeta_{n} \notag \\
  &+\frac{i}{8\alpha}\left[\phantom{\frac{1}{1}}\hspace{-.11in}G^{BC}\Omega^{mn}-4\alpha\left(T^{mn}\right)^{BC}\right]\Gamma_{ABC}\zeta_{n},
\end{align}
\begin{align}\label{eq:GKSE2}
  0=\delta \chi^{mn}_{r}=&\left[\phantom{\frac{1}{1}}\hspace{-.1in}T^{mn}_{AB}T_{CDrs}-\frac{1}{\alpha}T^{mn}_{AB}G_{CD}\Omega_{rs}+\frac{1}{12}\left(\mathcal{D}^{E}S^{[m}_{r}\delta^{n]}_{s}+\mathcal{D}_{F}T^{mn F E}\Omega_{rs}\right)\varepsilon_{EABCD}\right] \notag \\
  &\times\Gamma^{ABCD}\zeta^{s}+\left[ \frac{5}{2\alpha}T^{mn}_{AB}G^{A}_{\phantom{A}C}\Omega_{rs}-4 T^{mn}_{AB}T^{A}_{\phantom{A}Crs}+2T^{mn}_{BC}S_{rs}- S_{p}^{[m}T^{n]p}_{BC}\Omega_{rs}\right. \notag \\
  &\left.-R_{BC r}^{\phantom{B}[m}\delta^{n]}_{s}+\frac{1}{2}\mathcal{D}_{a}T^{mn}_{DE}\Omega_{rs}\varepsilon^{ADE}_{\phantom{ADE}BC}\right]\Gamma^{BC}\zeta^{s}+\left[\frac{1}{\alpha}T^{mn}_{AB}G^{AB}\Omega_{rs}-2 T^{mn}_{AB}T^{AB}_{rs}\right. \notag \\
  &\left.-\frac{4}{15}D^{mn}_{rs}\right]\zeta^{s} - (\mathrm{traces}).
\end{align}
In the above formulae, $\psi$ and $\chi$ are fermion fields in $(2,0)$ conformal supergravity,
$\zeta$ is a five dimensional spinor parameterizing rigid supersymmetry
transformations, i.e. the generalized Killing spinor, while $\{G, \alpha\}$ are
descendants of the six dimensional frame field and are therefore fixed. The
remaining bosonic background fields of supergravity $\{T, S, R, D\}$ are unknown
at this stage and need to be solved for. For definitions of covariant
derivatives as well as explanation of the $(\mathrm{traces})$ factor in the last
line of \eqref{eq:GKSE2}, see equations $(2.3)$ and $(2.17)$ in \cite[v1]{Cordova:2013bea}.

The above system of equations has twofold consequences. First of all,
it is a system of partial differential equations for the generalized Killing
spinor $\zeta$. The number of independent solutions to this system then gives
the number of preserved rigid supersymmetries of the theory defined on $M_5$. At
the same time it also provides constraints allowing to fix the values of bosonic
background fields $\{T, S, R, D\}$.

Our strategy for solving the system of equations \eqref{eq:GKSE1} and
\eqref{eq:GKSE2} consists first of specializing \eqref{eq:GKSE1} to our specific
supergravity background $M_5=S^2\times\mathbb{R}^2\times I$.\footnote{Remember, that we performed a twist on $\Sigma$, so we may treat it as flat space. Furthermore, the 5d spinor will be taken in a special factorized form with the component along $\mathbb{R}^2$ assumed to be constant. See \eqref{eq:SpinorFactorization} and discussion below.} This amounts to
inserting concrete values for the supergravity fields $G$ and $\alpha$, which
emerged from reduction of the frame field in six dimensions. Then we solve this
equation. By doing so, we get the generalized Killing spinors as well as a
subset of supergravity background fields $\{T, S, R\}$, i.e. all of them except
for $D$. Only after finishing this first step we move to the second equation
\eqref{eq:GKSE2} and plug in all the previously obtained background fields. This
simplifies the computation, since as we will see some terms vanish. Finally, we solve for $D$ from such simplified version of \eqref{eq:GKSE2}.

\subsection{Form of GKSE \eqref{eq:GKSE1} for $M_5=S^2\times\mathbb{R}^2\times I$}
Let us specialize equation \eqref{eq:GKSE1} to the geometry under consideration
$M_5=S^2\times\mathbb{R}^2\times I$. We convert the frame index to a coordinate one and write
\begin{equation}\label{eq:GKSE1system}
\delta\psi_{\mu}^{m}=\{\delta\psi_{\theta}^{m},\delta\psi_{\phi}^{m},\delta\psi_{x}^{m},\delta\psi_{y}^{m},\delta\psi_{u}^{m}\}=0 
\end{equation}
as a system of equations for each coordinate component. To present this system in a concise form we need to introduce some notation first, that makes manifest the representation decomposition of the spinor on $M_5$. Since the spin group of $M_5$ factorizes as $Spin(5)\to Spin(2)_{\mathbb{R}^2}\times Spin(2)_{S^2}$, it is natural to assume the decomposition\footnote{Before twisting the indices $\{\underbrace{\alpha,\sigma}_{\mathrm{Lorentz}}\vert\underbrace{\hat{\alpha},\hat{\sigma}}_{\mathrm{R-symm.}}\}$ correspond to $\{\underbrace{\mathfrak{so}(2)_{\mathbb{R}^2},\mathfrak{so}(2)_{S^2}}_{\mathrm{Lorentz}}\vert\underbrace{\mathfrak{so}(2)_{\hat{\alpha}},\mathfrak{so}(2)_{\hat{\sigma}}}_{\mathrm{R-symm.}}\}$. Twisting restricts $\mathfrak{so}(2)_{\mathbb{R}^2}\otimes\mathfrak{so}(2)_{\hat{\alpha}}$ to its diagonal subalgebra $\mathfrak{so}(2)^\prime$, therefore after twisting $\alpha,\hat{\alpha}$ transform under this $\mathfrak{so}(2)^\prime$ while $\sigma,\hat{\sigma}$ do not participate in the twisting procedure.}
\begin{equation}\label{eq:SpinorFactorization}
 \zeta_{\Lambda}^{m}=\epsilon_{\alpha}^{\widehat{\alpha}}\otimes\eta_{\sigma}^{\widehat{\sigma}},
\end{equation}
where $\epsilon$ is a spinor on $\mathbb{R}^2$ while $\eta$ is a spinor on $S^2$. The crucial assumption that we make about the spinors $\epsilon$ and $\eta$ are
\begin{align}\label{eq:SpinorConstraint}
\epsilon=const, && \eta=\eta(\theta,\phi). 
\end{align}
Action of an operator $O$ on $\zeta_{\Lambda}^{m}$ (the action is intended on both the spinor index $\Lambda$ and the R-symmetry index $m$) will then be represented in shorthand notation as
\begin{equation}\label{eq:SpinNotation}
 O\cdot\zeta=\left(\left[A\otimes\widehat{A}\right]\cdot\epsilon\right)\bigotimes\left(\left\{X\otimes\widehat{X}\right\}\cdot\eta\right)\equiv\left[A\otimes\widehat{A}\right]\left\{X\otimes\widehat{X}\right\}\left(\epsilon\otimes\eta\right).
\end{equation}
So here we have two rules:
\begin{itemize}
 \item operators in brackets act on spinors on $\mathbb{R}^2$ while those in braces act on spinors on $S^2$,
 \item unhatted operators act on spinor indices while hatted operators act on R-symmetry indices.
\end{itemize}

Now we arrived at a point when we can write the system of equations \eqref{eq:GKSE1} in a compact form. Plugging in our ansatz for the bosonic supergravity background fields \eqref{eq:SUGRAansatz} (all components that are not listed in \eqref{eq:SUGRAansatz} vanish up to symmetry relations among different components that are implicitly assumed)
and substituting explicit expressions for gamma matrices summarized in \eqref{eq:GammaMatrices} we get the first GKSE in the form outlined in \eqref{eq:GKSE1system}, in particular the order of the equations is as indicated there 
\begin{align}
0&=\Bigg(\partial_\theta\Big[\mathds{1}\otimes\widehat{\mathds{1}}\Big]\Big\{\mathds{1}\otimes\widehat{\mathds{1}}\Big\}+R\bigg(s\Big[\mathds{1}\otimes\widehat{\mathds{1}}\Big]\Big\{\kappa^1\otimes\widehat{\kappa}\Big\}+\widetilde{s}\Big[\mathds{1}\otimes\widehat{\kappa}\Big]\Big\{\kappa^1\otimes\widehat{\mathds{1}}\Big\} \notag \\
&-T\Big[\kappa^2\otimes\widehat{\kappa}^2\Big]\Big\{\kappa^1\otimes\widehat{\kappa}\Big\}+\widetilde{T}\Big[\kappa^1\otimes\widehat{\kappa}^1\Big]\Big\{\kappa^1\otimes\widehat{\kappa}\Big\}+\widetilde{t}\Big[\kappa\otimes\widehat{\kappa}\Big]\Big\{\kappa^1\otimes\widehat{\kappa}\Big\}\bigg)\Bigg)\zeta, \label{eq:GKSE1theta} \\[2ex]
0&=\Bigg(\partial_\phi\Big[\mathds{1}\otimes\widehat{\mathds{1}}\Big]\Big\{\mathds{1}\otimes\widehat{\mathds{1}}\Big\}+\frac{i}{2}\cos\theta\Big[\mathds{1}\otimes\widehat{\mathds{1}}\Big]\Big\{\kappa\otimes\widehat{\mathds{1}}\Big\} \notag \\ &+R\sin\theta\bigg(s\Big[\mathds{1}\otimes\widehat{\mathds{1}}\Big]\Big\{\kappa^2\otimes\widehat{\kappa}\Big\}+\widetilde{s}\Big[\mathds{1}\otimes\widehat{\kappa}\Big]\Big\{\kappa^2\otimes\widehat{\mathds{1}}\Big\} \notag \\
&-T\Big[\kappa^2\otimes\widehat{\kappa}^2\Big]\Big\{\kappa^2\otimes\widehat{\kappa}\Big\}+\widetilde{T}\Big[\kappa^1\otimes\widehat{\kappa}^1\Big]\Big\{\kappa^2\otimes\widehat{\kappa}\Big\}+\widetilde{t}\Big[\kappa\otimes\widehat{\kappa}\Big]\Big\{\kappa^2\otimes\widehat{\kappa}\Big\}\bigg)\Bigg)\zeta \label{eq:GKSE1phi}, \\[2ex]
0&=\Bigg(iw\Big[\mathds{1}\otimes\widehat{\kappa}^1\Big]\Big\{\mathds{1}\otimes\widehat{\mathds{1}}\Big\}+s\Big[\kappa^1\otimes\widehat{\mathds{1}}\Big]\Big\{\kappa\otimes\widehat{\kappa}\Big\}+\widetilde{s}\Big[\kappa^1\otimes\widehat{\kappa}\Big]\Big\{\kappa\otimes\widehat{\mathds{1}}\Big\} \notag \\ &+t\Big[\kappa^1\otimes\widehat{\kappa}\Big]\Big\{\mathds{1}\otimes\widehat{\kappa}\Big\}+\widetilde{T}\Big[\mathds{1}\otimes\widehat{\kappa}^1\Big]\Big\{\kappa\otimes\widehat{\kappa}\Big\}\Bigg)\zeta, \label{eq:GKSE1x} \\[2ex]
0&=\Bigg(-i\widetilde{w}\Big[\mathds{1}\otimes\widehat{\kappa}^2\Big]\Big\{\mathds{1}\otimes\widehat{\mathds{1}}\Big\}+s\Big[\kappa^2\otimes\widehat{\mathds{1}}\Big]\Big\{\kappa\otimes\widehat{\kappa}\Big\}+\widetilde{s}\Big[\kappa^2\otimes\widehat{\kappa}\Big]\Big\{\kappa\otimes\widehat{\mathds{1}}\Big\} \notag \\ &+t\Big[\kappa^2\otimes\widehat{\kappa}\Big]\Big\{\mathds{1}\otimes\widehat{\kappa}\Big\}-T\Big[\mathds{1}\otimes\widehat{\kappa}^2\Big]\Big\{\kappa\otimes\widehat{\kappa}\Big\}\Bigg)\zeta, \label{eq:GKSE1y} \\[2ex]
0&=\Bigg(iv\Big[\mathds{1}\otimes\widehat{\mathds{1}}\Big]\Big\{\mathds{1}\otimes\widehat{\kappa}\Big\}+i\widetilde{v}\Big[\mathds{1}\otimes\widehat{\kappa}\Big]\Big\{\mathds{1}\otimes\widehat{\mathds{1}}\Big\}+s\Big[\kappa\otimes\widehat{\mathds{1}}\Big]\Big\{\kappa\otimes\widehat{\kappa}\Big\} \notag \\ &+\widetilde{s}\Big[\kappa\otimes\widehat{\kappa}\Big]\Big\{\kappa\otimes\widehat{\mathds{1}}\Big\}+t\Big[\kappa\otimes\widehat{\kappa}\Big]\Big\{\mathds{1}\otimes\widehat{\kappa}\Big\}+\widetilde{t}\Big[\mathds{1}\otimes\widehat{\kappa}\Big]\Big\{\kappa\otimes\widehat{\kappa}\Big\}\Bigg)\zeta. \label{eq:GKSE1u}
\end{align}
Note that \eqref{eq:GKSE1theta} and \eqref{eq:GKSE1phi} are partial differential equations for the spinor $\eta$ on $S^2$, while \eqref{eq:GKSE1x} -- \eqref{eq:GKSE1u} are constraints for the supergravity background fields and for the spinor $\epsilon$ on $\mathbb{R}^2$.

To proceed we take a linear combination\footnote{We thank the referee for pointing this out to us, making thus the solution more compact.} $\big[\{\kappa\}\eqref{eq:GKSE1phi}-i\sin\theta\eqref{eq:GKSE1theta}\big]$ resulting in
\begin{align}\label{eq:GKSEcomb}
\Big[\big\{\kappa\otimes\widehat{\mathds{1}}\big\}\partial_\phi+\big\{\mathds{1}\otimes\widehat{\mathds{1}}\big\}\left(\frac{i}{2}\cos\theta-i\sin\theta\partial_\theta\right)\Big]\eta=0.
\end{align}
The differential operator acts as identity on the R-symmetry index and produces thus two identical equations  for two columns of the matrix $\eta$ labeled by R-symmetry. We pick one and call it $\widetilde{\eta}$. Let us choose anti-periodic boundary conditions under $\phi\to\phi+2\pi$ for $\widetilde{\eta}$, i.e. we are working in the Neveu--Schwarz sector. Then the Fourier expansion has the form
\begin{equation}
\widetilde{\eta}=\sum_{m\in\mathbb{Z}+\frac{1}{2}}c_m(\theta)e^{im\phi}
\end{equation}
and from \eqref{eq:GKSEcomb} we get ordinary differential equations for the coefficients $c_m(\theta)$
\begin{align}
\frac{d}{d\theta}\begin{pmatrix}c_m^+(\theta) \\ c_m^-(\theta) \end{pmatrix}=\begin{pmatrix}\lambda_m^+(\theta) & 0 \\ 0 & \lambda_m^-(\theta)\end{pmatrix}\begin{pmatrix}c_m^+(\theta) \\ c_m^-(\theta) \end{pmatrix}
\end{align}
where
\begin{equation}
\lambda_m^{\pm}(\theta)=\left(\pm m-\frac{1}{2}\right)\frac{1}{2}\tan\frac{\theta}{2}+\left(\pm m+\frac{1}{2}\right)\frac{1}{2}\cot\frac{\theta}{2}.
\end{equation}
The solution to this system of equations reads
\begin{align}
\widetilde{\eta}=\sum_{m\in\mathbb{Z}+\frac{1}{2}}\begin{pmatrix}K_m^+ &\left(\sin\frac{\theta}{2}\right)^{m+\frac{1}{2}}&&\left(\cos\frac{\theta}{2}\right)^{-m+\frac{1}{2}} \\ K_m^- &\left(\sin\frac{\theta}{2}\right)^{-m+\frac{1}{2}}&&\left(\cos\frac{\theta}{2}\right)^{m+\frac{1}{2}}\end{pmatrix}e^{im\phi}.
\end{align}
Requiring regularity forces us to restrict the range of modes to $m=\pm\frac{1}{2}$ only (otherwise vanishing denominators appear at poles of $S^2$).

We have a linear four-dimensional space of solutions spanned by $\{K_{+\frac{1}{2}}^+,K_{+\frac{1}{2}}^-,K_{-\frac{1}{2}}^+,K_{-\frac{1}{2}}^-\}$. Therefore the final expression for $\eta$ can be written in the form
\begin{equation}
\eta=\beta^{ij}\eta_{ij};\quad i,j=\pm, 
\end{equation}
where $\beta^{ij}$ are constants and $\eta_{ij}$ is a basis of solutions (rows are labeled by spinor index while columns by R-symmetry index)
\begin{align}
\eta_{++}&=
\begin{bmatrix}
0 & \sin\frac{\theta}{2}e^{\frac{i}{2}\phi} \\
0 & \cos\frac{\theta}{2}e^{\frac{i}{2}\phi}
\end{bmatrix},
&
\eta_{+-}&=
\begin{bmatrix}
-\sin\frac{\theta}{2}e^{\frac{i}{2}\phi} & 0 \\
\cos\frac{\theta}{2}e^{\frac{i}{2}\phi} & 0
\end{bmatrix},
\notag \\[2ex]
\eta_{-+}&=
\begin{bmatrix}
0 & -\cos\frac{\theta}{2}e^{-\frac{i}{2}\phi} \\
0 & \sin\frac{\theta}{2}e^{-\frac{i}{2}\phi}
\end{bmatrix},
&
\eta_{--}&=
\begin{bmatrix}
\cos\frac{\theta}{2}e^{-\frac{i}{2}\phi} & 0 \\
\sin\frac{\theta}{2}e^{-\frac{i}{2}\phi} & 0
\end{bmatrix}.
\end{align}
The columns of the above matrices form individual Killing spinors and so we see four of them being independent. Let us group them as follows
\begin{align}
\left. \eta_1=e^{-\frac{i}{2}\phi}\begin{pmatrix}\cos\frac{\theta}{2} \\ \sin\frac{\theta}{2}\end{pmatrix},\;
       \overline{\eta}_1=e^{\frac{i}{2}\phi}\begin{pmatrix}\sin\frac{\theta}{2} \\ \cos\frac{\theta}{2}\end{pmatrix} \quad\right|\quad 
       \eta_2=e^{-\frac{i}{2}\phi}\begin{pmatrix}-\cos\frac{\theta}{2} \\ \sin\frac{\theta}{2}\end{pmatrix},\;
       \overline{\eta}_2=e^{\frac{i}{2}\phi}\begin{pmatrix}-\sin\frac{\theta}{2} \\ \cos\frac{\theta}{2}\end{pmatrix}
\end{align}
in order to facilitate comparison with \cite{Doroud:2012xw} or \cite{Benini:2012ui}.
Indeed, $(\eta_1,\overline{\eta}_1)$ and  $(\eta_2,\overline{\eta}_2)$ precisely agree with the two pairs of conformal Killing spinors given in \cite{Doroud:2012xw} parameterizing $\mathcal{N}=(2,2)$ superconformal symmetry on $S^2$. 

Now we move to fixing the supergravity background fields using the constraints \eqref{eq:GKSE1x} -- \eqref{eq:GKSE1u}. The matrices in braces on the right hand side of these equations act on the nontrivial spinor $\eta$, producing thus independent images of it. We require that the coefficient of each such term vanishes separately, which fixes the supergravity background fields and moreover imposes a constraint on the spinor $\epsilon$. This leads to
\begin{align}
\eqref{eq:GKSE1x} &\Longrightarrow && w=t=\widetilde{s}=0 &&& s\big[\kappa^1\otimes\widehat{\mathds{1}}\big]+\widetilde{T}\big[\mathds{1}\otimes\widehat{\kappa}^1\big]=0 \notag \\
\eqref{eq:GKSE1y} &\Longrightarrow && \widetilde{w}=t=\widetilde{s}=0 &&& s\big[\kappa^2\otimes\widehat{\mathds{1}}\big]-T\big[\mathds{1}\otimes\widehat{\kappa}^2\big]=0 \notag \\
\eqref{eq:GKSE1u} &\Longrightarrow && v=\widetilde{v}=t=\widetilde{s}=0 &&& s\big[\kappa\otimes\widehat{\mathds{1}}\big]+\widetilde{t}\big[\mathds{1}\otimes\widehat{\kappa}\big]=0
\end{align}
From the last column above we see that in order to find a solution ($s=\widetilde{t}=T=\widetilde{T}=0$ is not a consistent solution) we need $\epsilon$ to be antisymmetric. Then the factors in the tensor product can be swapped with a change of sign and \eqref{eq:GKSE1x} -- \eqref{eq:GKSE1u} are solved by setting
\begin{equation}
\tau\equiv T=-\widetilde{T}=-\widetilde{t}=-s.
\end{equation}
To fix $\tau$ we come back to \eqref{eq:GKSE1theta} and \eqref{eq:GKSE1phi}. After substituting the values for supergravity background fields found so far they can be presented in the form (here $\eta_{\pm}$ means the R-symmetry doublet of spinors forming the columns of $\eta$)
\begin{align}\label{eq:KillSpS2}
\nabla_z\eta_+=2\tau\gamma_z^{(2D)}\eta_+,\qquad \nabla_z\eta_-=-2\tau\gamma_z^{(2D)}\eta_-. 
\end{align}
Taking the commutator of either of these equations yields the integrability condition
\begin{equation}
 \frac{1}{4}R_{zw\bar{z}\bar{w}}\gamma_{(2D)}^{\bar{z}}\gamma_{(2D)}^{\bar{w}}=-4\tau^2\left[\gamma^{(2D)}_z,\gamma^{(2D)}_w\right].
\end{equation}
Multiplying by $\gamma_{(2D)}^z\gamma_{(2D)}^w$ from the left results in a relation between $\tau$ and the scalar curvature of $S^2$
\begin{equation}
-\frac{R^{(S^2)}_{\mathrm{SC}}}{2}=16\tau^2,
\end{equation}
which immediately gives $\tau=\pm\tfrac{i}{4R}$. We will pick the plus sign in the following.

From 5d point of view, the  spinor \eqref{eq:SpinorFactorization} is the product of spinor $\epsilon$ on $\mathbb{R}^2$ and $\eta$ on $S^2$, which have four and eight components, respectively. Therefore, there are 32 supercharges. (Note that the 6d (2,0) SCFT has 16 supercharges which can be extended to 32 supercharges forming the superconformal algebra explicitly, and the reduction on $S^1$ to 5d SYM should not break these supercharges. This is consistent with the number of the supercharges we get here.) After we consider general manifold $\Sigma$, we need to perform the topological twist on $\Sigma$ discussed in section \ref{S:geometry}, 
which turns two components of the spinor $\epsilon$ into two scalars. Hence 16 supercharges -- which are scalars on $\Sigma$ -- are preserved in the 5d theory.

\subsection{Solving for the $D$ background field from GKSE}
It is an appropriate place here to summarize the list of bosonic supergravity background fields, which were obtained in previous subsection and are needed to simplify the second GKSE \eqref{eq:GKSE2}. The table \eqref{eq:SUGRAansatz} completes to
\begin{align}\label{eq:SUGRAbackground}
G=0, &&
\alpha=\widetilde{R}^{-1}, && 
S_{\hat{A}\hat{B}}=-\frac{i}{4R}\varepsilon_{\hat{x}\hat{y}}, &&
V_{A\hat{B}\hat{C}}=0, &&
T_{\hat{A}BC}=-\frac{i}{4R}\varepsilon_{\hat{a}bc}.
\end{align}
In such a background \eqref{eq:GKSE2} simplifies: terms that manifestly vanish are those containing $G$, the field strength $R$ for the R-symmetry gauge field $V$ and also terms with covariant derivatives, either $\mathcal{D}T$ or $\mathcal{D}S$. Without using yet the explicit form for $T$ and $S$ the simplified version reads
\begin{align}\label{eq:GKSE2simp}
0=\delta \chi^{mn}_{r}=&\phantom{\frac{1}{1}}\hspace{-.1in}T^{mn}_{AB}T_{CDrs}\Gamma^{ABCD}\zeta^{s} 
                         +\left[-4 T^{mn}_{AB}T^{A}_{\phantom{a}Crs}+2T^{mn}_{BC}S_{rs}- S_{p}^{[m}T^{n]p}_{BC}\Omega_{rs}\right]\Gamma^{BC}\zeta^{s} \notag \\
                         &+\left[-2 T^{mn}_{AB}T^{AB}_{rs}-\frac{4}{15}D^{mn}_{rs}\right]\zeta^{s}-(\mathrm{traces}).
\end{align}
It is straightforward to show that the first term vanishes, indeed the $T_{AB}T_{CD}$ term has at least one pair of common frame indices from the set $\{3,4,5\}$ while $\Gamma^{ABCD}$ is completely anti-symmetric. Somewhat more involved is to see that the last term in first line vanishes as well. We have
\begin{align}
\bigg[S_p^mT_{BC}^{np}-(m\leftrightarrow n)\bigg]&=\bigg[\left(-\frac{i}{4R}\right)\varepsilon_{\hat{x}\hat{y}}\left(\Gamma^{\hat{x}\hat{y}}\right)_p^m\left(-\frac{i}{4R}\right)\varepsilon_{\hat{d}bc}\left(\Gamma^{\hat{d}}\right)^{np}-(m\leftrightarrow n)\bigg] \notag \\ 
&=-\frac{1}{(4R)^2}\varepsilon_{\hat{x}\hat{y}}\varepsilon_{\hat{d}bc}\Big[\Gamma^{\hat{d}},\Gamma^{\hat{x}\hat{y}}\Big]
\end{align}
and the commutator $\Big[\Gamma^{\hat{d}},\Gamma^{\hat{x}\hat{y}}\Big]$ vanishes, since $\hat{d}\in\{3,4,5\}$ while $\hat{x},\hat{y}\in\{1,2\}$. Therefore the only surviving terms in \eqref{eq:GKSE2simp} are
\begin{align}\label{eq:GKSE2Fin}
0=\delta \chi^{mn}_{r}=&\bigg[\underbrace{-4 T^{mn}_{AB}T^{A}_{\phantom{A}Crs}}_{\circled{1}}+\underbrace{2T^{mn}_{BC}S_{rs}}_{\circled{2}}\bigg]\Gamma^{BC}\zeta^{s}+\bigg[\underbrace{-2 T^{mn}_{AB}T^{AB}_{rs}}_{\circled{3}}-\frac{4}{15}D^{mn}_{rs}\bigg]\zeta^{s}-(\mathrm{traces}). 
\end{align}

Now, we are going to provide the (traces) part for the individual numbered terms (the $D$ field is already traceless). For some examples of these computations see (A.9) in \cite[v2]{Bergshoeff:1999db}. The results of this little calculation are
\begin{align}
&\circled{1}: && -4\Gamma^{BC}\bigg\{T_{AB}^{mn}T^{A}_{\phantom{A}Crs}-\frac{4}{5}\delta^{[n}_rT_{AB}^{m]q}T^A_{\phantom{A}Cqs}+\frac{1}{5}\Omega^{mn}T_{ABr}^{\phantom{ABr}q}T^A_{\phantom{A}Cqs}\bigg\} \label{eq:T1}
\\[1.5em]
&\circled{2}: && 2\Gamma^{BC}\bigg\{T_{BC}^{mn}S_{rs}-\frac{4}{5}\delta^{[n}_rT_{BC}^{m]q}S_{qs}+\frac{1}{5}\Omega^{mn}T_{BCr}^{\phantom{BCr}q}S_{qs}\bigg\} \label{eq:T2}
\\[1.5em]
&\circled{3}: && -2\bigg\{T_{AB}^{mn}T_{rs}^{AB}-\frac{4}{5}\delta^{[n}_rT_{AB}^{m]q}T_{qs}^{AB}+\frac{1}{5}\Omega^{mn}T_{ABr}^{\phantom{ABr}q}T^{AB}_{qs}\bigg\} \label{eq:T3}.
\end{align}
The ansatz proposed for the $D$ field has the form
\begin{equation}
D_{\hat{A}\hat{B}}=\mathrm{diag}\left(d_1,d_1;d_2,d_2;-2(d_1+d_2)\right),
\end{equation}
which in turn induces
\begin{align}\label{eq:Dansatz}
D^{mn}_{rs}=&d_1\bigg[\left(\Gamma^{\hat{1}}\right)^{mn}\left(\Gamma^{\hat{1}}\right)_{rs}+\left(\Gamma^{\hat{2}}\right)^{mn}\left(\Gamma^{\hat{2}}\right)_{rs}\bigg]+d_2\bigg[\left(\Gamma^{\hat{3}}\right)^{mn}\left(\Gamma^{\hat{3}}\right)_{rs}+\left(\Gamma^{\hat{4}}\right)^{mn}\left(\Gamma^{\hat{4}}\right)_{rs}\bigg] \notag \\
&-2(d_1+d_2)\left(\Gamma^{\hat{5}}\right)^{mn}\left(\Gamma^{\hat{5}}\right)_{rs}.
\end{align}

Substituting \eqref{eq:T1}--\eqref{eq:T3} together with \eqref{eq:Dansatz} into \eqref{eq:GKSE2Fin} gives the final matrix equation needed to fix the unknown coefficients $d_1$ and $d_2$ entering the ansatz for the $D$ field \eqref{eq:Dansatz}. Perhaps the most efficient way how to get and solve a system of equations for these two coefficients is to use a computer algebra software. We performed the computation in \texttt{Mathematica} with the result
\begin{align}
d_1=-\frac{9}{16R^2}, && d_2=\frac{3}{8R^2}.
\end{align}

\section{Reduction to 3d Theory}\label{S:reduction}
The 5d Super-Yang-Mills action consists of four terms
\begin{equation}\label{eq:fullaction}
S = S_{A} + S_{\phi} + S_{\rho} + S_{int}.
\end{equation}
They are constructed by substituting the bosonic background fields given in \eqref{eq:SUGRAbackground} into the general formulae \eqref{eq:SA}--\eqref{eq:Sint}. To keep compact notation we switch between $\mathfrak{so}(5)/\mathfrak{sp}(4)$ notation as appropriate. (Note that in the following we define $\phi_{mn}=\phi_{\widehat{A}}\big(\Gamma^{\widehat{A}}\big)_{mn}$. And our conventions for the indices and gamma matrices are summarized in Appendix A and B.) We list the four terms below:
\begin{align}
S_A&=\frac{1}{8\pi^2\widetilde{R}}\int_{M_5}\mathrm{Tr}(F\wedge\star F), \label{eq:SAsub} \\
S_\phi&=\frac{1}{8\pi^2\widetilde{R}}\int_{M_{5}}d^5x\sqrt{g}~\mathrm{Tr}\left(\mathcal{D}_\mu\phi_{\widehat{A}}\mathcal{D}^\mu\phi^{\widehat{A}}-\frac{i}{R}\epsilon^{\hat{a}bc}\phi_{\hat{a}}F_{bc}\right) \label{eq:Sphisub}, \\
S_\rho&=\frac{i}{32\pi^2\widetilde{R}}\int_{M_5}d^5x\sqrt{g}~\mathrm{Tr}\left(\rho_m\slashed{\mathcal{D}}\rho^m-\frac{1}{8R}\rho_m\left[\epsilon_{\hat{z}\hat{w}}\left(\Gamma^{\hat{z}\hat{w}}\right)^{mn}-\epsilon_{\hat{a}bc}\left(\Gamma^{\hat{a}}\right)^{mn}\Gamma^{bc}\right]\rho_n\right), \\
S_{int}&=\frac{1}{32\pi^2\widetilde{R}}\int_{M_5}d^5x\sqrt{g}~\mathrm{Tr}\left(\rho_m\left[\phi^{mn},\rho_n\right]-\frac{1}{4}\left[\phi_{mn},\phi^{nr}\right]\left[\phi_{rs},\phi^{sm}\right]+i\frac{4}{3R}\epsilon^{abc}\phi_a\left[\phi_b,\phi_c\right]\right) \label{eq:Sintsub}.
\end{align}
Spinor indices are implicit in these expressions and the covariant derivatives are defined as
\begin{align}
\mathcal{D}_\mu\phi_{\widehat{A}}&=\partial_\mu\phi_{\widehat{A}}+\left[A_\mu,\phi_{\widehat{A}}\right], \\
\mathcal{D}_\mu\rho^m&=\left(\partial_\mu+\frac{1}{4}\omega_\mu^{AB}\Gamma_{AB}\right)\rho^m+\left[A_\mu,\rho^m\right].
\end{align}
Observe that mass terms for the scalar fields $\phi_{\widehat{A}}$ induced by supergravity canceled among each other, leaving thus all these fields massless in the 5d SYM. 

With the five-dimensional action we obtained, we are now proceeding to do the reduction on the sphere. We will see that after the reduction, the massless modes of the fields on the sphere give rise to the complex Chern-Simons theory.  
Let us analyze the four terms' reduction in \eqref{eq:fullaction} to three dimensions respectively.

\subsection{Gauge fields}
The purpose of this section is to study the structure of massless modes of the gauge field $A$, which emerges after dimensional reduction on $S^2$. The result we are going to establish here is
\begin{align}\label{eq:Amassmodestruct}
\textrm{zero modes of }A_b:\quad&\textrm{massless} \notag \\
\textrm{higher modes of }A_b:\quad&\textrm{mass }\propto\frac{1}{R}\;\Rightarrow\;\textrm{decoupled} \notag \\
\textrm{all modes of }A_{\Theta}:\quad&\textrm{mass }\propto\frac{1}{R}\;\Rightarrow\;\textrm{decoupled}, \notag \\
\end{align}
where $b$ denote the indices on $\Sigma \times I$~\footnote{Note that after the topological twist, $\Sigma$ is taken to be flat and thus its coframe vector indices coincide with the curved vector indices.} and $\Theta = \{\theta, \phi\}$ denote the Lorentz vector indices on $S^2$, and the precise definition of the zero modes will be given in a moment. The proof of this statement will be based on harmonic analysis on $S^2$.

Effective masses for Kaluza--Klein modes of the gauge field originate purely from the five dimensional Yang-Mills kinetic term, so we may concentrate just on this piece of the whole action
\begin{align}\label{eq:Akin}
\frac{1}{8\pi^2\tilde{R}}\int_{M_5}\mathrm{Tr}\left(F\wedge\star F\right)&=\frac{1}{16\pi^2\widetilde{R}}\int_{M_5} d^5x\sqrt{g}~\mathrm{Tr}F_{\mu\nu}F^{\mu\nu} \notag \\
&=-\frac{1}{8\pi^2\widetilde{R}}\int_{M_5} d^5x\sqrt{g}~\mathrm{Tr}A_\mu\left[\nabla_\rho\nabla^\rho g^{\mu\nu}-\mathrm{Ricc}^{\mu\nu}\right]A_{\nu} \notag \\ &\ \ \ +\left(\textrm{cubic and quartic interaction terms for non-abelian YM}\right).
\end{align}
Here we used the gauge fixing condition $\nabla_\mu A^\mu=0$ together with careful integration by parts to arrive at the second line, and as discussed above extracted just the kinetic term. The operator in brackets will generate the effective masses after dimensional reduction on $S^2$, hence let us focus our attention to it:
\begin{align}\label{eq:LapDecomp}
A_\mu\left[\nabla_\rho\nabla^\rho g^{\mu\nu}-\mathrm{Ricc}^{\mu\nu}\right]A_{\nu}&=A_b\Delta_{M_5}A^b+A_\Theta\left(\Delta_{M_5}-\frac{1}{R^2}\right)A^\Theta \notag \\
&=A_b\left(\Delta_{M_3}+\left\{\Delta_{S^2}\right\}\right)A^b+A_\Theta\left(\Delta_{M_3}+\left\{\Delta^{(H);1}_{S^2}\right\}\right)A^\Theta.
\end{align}
In the above equation we split the Laplace--Beltrami operator $\Delta_{M_5}$ in accordance with dimensional reduction on $S^2$, i.e. into $\Delta_{M_3}$ leading to kinetic terms in the effective action on $M_3=\Sigma\times I$ and into $\Delta_{S^2}$ producing effective mass terms (highlighted by braces). Here $\Delta_{S^2}^{(H);1}$ is the Hodge Laplacian (Laplace--de Rham operator) on one-forms and its relation to the connection Laplacian (Laplace--Beltrami operator) is given as
\begin{align}
\Delta_{S^2}^{(H);1}=\Delta_{S^2}-\frac{1}{R^2}.
\end{align}

Now, we expand individual components of the gauge connection to eigenmodes of the operators denoted by braces in \eqref{eq:LapDecomp}, i.e. harmonic functions on $S^2$
\begin{align}
A_b(M_5)=\sum_{l\geq 0}\sum_{\vert m\vert\leq l}A_b^{l,m}(M_3)Y^{l,m}(S^2) \\
A_\Theta(M_5)=\sum_{l\geq 1}\sum_{\vert m\vert\leq l}\varphi^{l,m}(M_3)\mathcal{Y}_\Theta^{l,m}(S^2),
\end{align}
where $A_b^{l,m}$ (one-forms on $M_3$) and $\varphi^{l,m}$ (scalars on $M_3$) represent Kaluza--Klein modes of the five dimensional gauge connection while $Y^{l,m}$ and $\mathcal{Y}_\Theta^{l,m}$ are scalar and vector spherical harmonics, respectively. Their spectrum with respect to the Laplace--Beltrami operator on $S^2$ takes the form \cite{Chodos:1983zi}(Sec. III)
\begin{align}
\Delta_{S^2}Y^{l,m}=\lambda_l^{(S)}Y^{l,m};\quad &\lambda_l^{(S)}=-\frac{1}{R^2}l(l+1),\quad l=0,1,2,\ldots \notag \\
						 &d_{\lambda^{(S)}}(l)=2l+1 \label{eq:specS}\\
\Delta_{S^2}\mathcal{Y}_\Theta^{l,m}=\lambda_l^{(V)}\mathcal{Y}_\Theta^{l,m};\quad &\lambda_l^{(V)}=\left\{
\begin{array}{lr}
-\frac{1}{R^2}\left[l(l+1)-1\right]\;\textrm{(transverse)} \\
-\frac{1}{R^2}\left[l(l+1)-1\right]\;\textrm{(longitudinal)}
\end{array}\right\}
\begin{split}
&d_{\lambda^{(V)}}=2l+1 \\ &l=1,2,\ldots . 
\end{split}
\label{eq:specV}
\end{align}
In these expressions $d_\lambda(l)$ denotes the multiplicity of the corresponding eigenvalue $\lambda$ and note that for vector harmonics the case $l=0$ is omitted.

Using orthonormality of spherical harmonics together with formulae \eqref{eq:specS}, \eqref{eq:specV} one can easily integrate the kinetic term in \eqref{eq:Akin} over $S^2$ to get the effective action for Kaluza--Klein modes
\begin{align}
&\sum_{l\geq0}\sum_{\vert m\vert\leq l}\int_{M_3}A_b^{l,m}(M_3)\left[\Delta_{M_3}-\left\{\frac{l(l+1)}{R^2}\right\}\right]A^b_{l,m}(M_3) \notag \\
+& \sum_{l\geq1}\sum_{\vert m\vert\leq l}\int_{M_3}\varphi^{l,m}(M_3)\left[\Delta_{M_3}-\left\{\frac{l(l+1)}{R^2}\right\}\right]\varphi^{l,m}(M_3).
\end{align}
Therefore we can conclude that the scalar Kaluza--Klein modes $\varphi^{l,m}$ (originating from components on $S^2$ of the five dimensional connection) all decouple, since their masses are proportional to $\tfrac{1}{R}$ and tend to infinity in the reduction limit $R\to 0$~\footnote{Indeed, this was evident already from \eqref{eq:LapDecomp}, since the operator producing effective masses for $\varphi^{l,m}$ is the Hodge Laplacian $\Delta_{S^2}^{(H);1}$. However, it is a well known result that $b^1(S^2)=0$, i.e. there are no harmonic one forms,  $\Delta_{S^2}^{(H);1}A_\Theta=0$, on the sphere and thus all modes must be massive with the mass scale set to $\tfrac{1}{R}$.}. The same is true for all $l>0$ modes $A_b^{l,m}$. There is a single massless mode $A_b^{0,0}(M_3)$, which was referred to as the zero mode in \eqref{eq:Amassmodestruct}.

With this analysis at our disposal it is straightforward to perform the dimensional reduction on $S^2$ in \eqref{eq:Akin}. The final outcome is the effective action for this single massless mode (the superscript $(0,0)$ will be dropped from now on)
\begin{align}
S_A=\frac{R^2}{4\pi\widetilde{R}}\int_{M_3}d^3x\sqrt{g_{M_3}}~\mathrm{Tr}(F_{ab}F^{ab}).
\end{align}

\subsection{Scalar fields}
The plan of the reduction process is similar as for the gauge fields, thus we will be brief. As a first step the kinetic term has to be isolated from \eqref{eq:Sphisub}. Afterward one integrates by parts to obtain
\begin{align}
\phi_{\widehat{A}}\left(\Delta_{M_3}+\left\{\Delta_{S^2}\right\}\right)\phi^{\widehat{A}},
\end{align}
where $\Delta_{M_3}$ leads to kinetic terms in the effective theory on $M_3$, while $\Delta_{S^2}$ assigns effective masses to the Kaluza--Klein modes. Again, we may expand $\phi_{\widehat{A}}$ into scalar spherical harmonics, which are eigenfunctions of $\Delta_{S^2}$ with eigenvalues $-\tfrac{1}{R^2}l(l+1),\;l\geq 0$. Their multiplicity is $2l+1$, hence we see that there is a single massless mode corresponding to $l=0$, i.e. a constant zero mode. All other modes have masses proportional to $\tfrac{1}{R}$, thus become infinitely massive in the limit $R\to 0$ and decouple from the low energy effective theory.

Keeping only the massless modes collected so far, we compute the low energy effective action by integrating \eqref{eq:Sphisub} over $S^2$. Since all these modes are constant on the sphere it is a straightforward calculation leading to
\begin{align}
S_\phi=\frac{R^2}{2\pi\widetilde{R}}\int_{M_3}d^3x\sqrt{g_{M_3}}~\mathrm{Tr}\left(\mathcal{D}_b\phi_{\hat{z}}\mathcal{D}^b\phi^{\hat{z}}+\mathcal{D}_b\phi_{\hat{a}}\mathcal{D}^b\phi^{\hat{a}}-\frac{i}{R}\epsilon^{\hat{a}bc}\phi_{\hat{a}}F_{bc}\right). 
\end{align}

\subsection{Fermions}
In the five-dimensional Super-Yang-Mills theory, the kinetic and mass action for the fermions is 
\begin{equation}
S_\rho = \frac{1}{32\pi^2\tilde{R}} \int_{S^2 \times M_3} d^5x\sqrt{g}\ \text{Tr} \left( \rho_{m\Lambda} i \slashed{D}^{\Lambda}_{\Pi} \rho^{m\Pi} + \rho_{m\Lambda} M^{mn\Lambda}_{\rho\Pi} \rho_{n}^{\Pi}\right).
\end{equation}
Here we have 
 \[\slashed{D} = \Gamma^\mu (\nabla_\mu + [A_\mu, \bullet]),\]
\[\quad M_\rho^{mn} = \frac{1}{2} S^{mn} - \frac{1}{2} \slashed{T}^{mn} = -\frac{1}{8R}\left[\epsilon_{\hat{z}\hat{w}}\left(\Gamma^{\hat{z}\hat{w}}\right)^{mn}-\epsilon_{\hat{a}bc}\left(\Gamma^{\hat{a}}\right)^{mn}\Gamma^{bc}\right],\]
where the background fields $S$ and $T$ defined in Appendix \ref{conventions} are solved in Section \ref{S:geometry} and \ref{S:GKSE}. And note that with all the massive modes decoupled, we have the gauge field components on the sphere $A_z = 0$, which leads to 
\begin{equation}
\slashed{D}_{S^2} = \slashed{\nabla}_{S^2} = \Gamma^z \nabla_z.
\end{equation}

In order to do the dimensional reduction on $S^2$, we need to decompose the fermions according to their profile on the sphere. To do so, we first decompose the fermions with respect to their representation under the symmetry group $SO(2)_R \times SO(2)_R \times SO(2)_L \times SO(2)_L$\footnote{
Here in the reduction down to the three-dimensional theory on $I \times \Sigma$, we actually perform the decomposition under the symmetry group $SO(2)_R \times SO(3)_R \times SO(2)_L \times SO(3)_L$ following the prescription in \cite{Cordova:2013bea}. This is because the Lorentz group is a priori $SO(3)$  for $M_3 = I\times\Sigma$. However, due to the boundaries of $M_3$ -- on which we will obtain a complex Toda theory after furture reduction -- the symmetry group $SO(3)$ reduces to $SO(2)$.
}: 
\begin{equation}\label{fexp}
\rho^{m\Lambda} = \epsilon^{\alpha \hat{\alpha}} \lambda^{\sigma \hat{\sigma}} + (\kappa^{a})^{\alpha \hat{\alpha}} \xi_a^{\sigma \hat{\sigma}},
\end{equation}
where $\lambda^{\sigma \hat{\sigma}} $ and $\xi_a^{\sigma \hat{\sigma}}$ are Grassmann fields dependent on all coordinates of $M_5$ manifestly taking the indices of the symmetry group $SO(2)_L\times SO(2)_R$ (attached to $S^2$). After the topological twist, $\lambda^{\sigma \hat{\sigma}}$ transform as scalars under the left over unbroken diagonal subgroup of the complementary $SO(2)_R \times SO(2)_L$, while $\xi_a^{\sigma \hat{\sigma}}$ as vectors on $\Sigma\times I$. The next step is to take the modes decomposition on the sphere. Here following \cite{Cordova:2013cea}, we choose a convenient basis of modes which are eigenfunctions of the operator $\kappa \slashed{\nabla}_{S^2}$: 
\begin{equation}\label{eigen}
\left(\kappa\cdot\slashed{\nabla}_{S^2}\right)_\tau^\sigma \Theta^{\tau} = \frac{n}{R} \Theta^\sigma,\ \  n \in \mathbb{Z}_{\slashed{=}0},
\end{equation}
with $\kappa$ defined in Appendix \ref{App:GammaMat}. 
The modes satisfy the orthogonality condition: 
\begin{equation}
\int_{S^2} d^2x \sqrt{g} \Theta^{\sigma}\tilde{\Theta}^{\tau}B_{\sigma\tau} \propto \frac{1}{R}\delta(n + \tilde{n} ),\ \ \int_{S^2} d^2x \sqrt{g} \Theta^{\sigma}\tilde{\Theta}^{\tau}\kappa_{\sigma\tau} \propto \frac{1}{R}\delta(n - \tilde{n} ). 
\end{equation} 

To proceed the reduction, let's first consider taking $n=\pm 1$ and defining the associated modes as\footnote{To be specific, the modes $a,b$ satisfy a stronger equation than \eqref{2deigen}. They are Killing spinors on the punctured sphere (to avoid multi-valuedness at poles), related to \eqref{eq:KillSpS2} by appearance of the chirality matrix $\kappa$ on the right-hand sides. Their normalization is chosen for later convenience. Being Killing spinors is fully equivalent to \eqref{2deigen} together with the additional requirement of being conformal Killing spinors (known as twistor spinors in math literature). In equations 
\begin{align}
\left.\begin{array}{l}
\kappa\slashed{\nabla}a=\frac{1}{R}a \\ \nabla_z a-\frac{1}{2}\gamma_z^{(2D)}\slashed{D}a=0
\end{array}\right\}
\Leftrightarrow \nabla_z a=\frac{1}{2R}\gamma_z^{(2D)}\kappa a;
&&
\left.\begin{array}{l}
\kappa\slashed{\nabla}b=-\frac{1}{R}b \\ \nabla_z b-\frac{1}{2}\gamma_z^{(2D)}\slashed{D}b=0
\end{array}\right\}
\Leftrightarrow \nabla_z b=-\frac{1}{2R}\gamma_z^{(2D)}\kappa b. \notag
\end{align}
} $a^{\sigma}_i$ and $b^{\sigma}_i$: 
\begin{equation}\label{2deigen}
\left(\kappa\cdot \slashed{\nabla}_{S^2}\right)^{\sigma}_{\tau} a^{\tau}_{i} =   \frac{1}{R} a^{\sigma}_{i}, \ \  \left(\kappa\cdot \slashed{\nabla}_{S^2}\right)^{\sigma}_{\tau} b^{\tau}_{i} =   -\frac{1}{R} b^{\sigma}_{i},
\end{equation}
where $i = \pm$ (also denoted $i=1,\,2$) label two types of distinct solutions to the above equations. Explicitly, they take the following form: 
\[ a^{\sigma}_{+} = \frac{e^{i\phi/2}}{\sqrt{4\pi R}} \left[ \begin{array}{cc}
i \sin(\theta/2) \\
\cos(\theta/2)  \end{array} \right],  \ \ 
a^{\sigma}_{-} = \frac{e^{-i\phi/2}}{\sqrt{4\pi R}} \left[ \begin{array}{cc}
\cos(\theta/2) \\
i \sin(\theta/2)   \end{array} \right],\]

\[ b^{\sigma}_{+} = \frac{e^{i\phi/2}}{\sqrt{4\pi R}} \left[ \begin{array}{cc}
-i \sin(\theta/2) \\
\cos(\theta/2)  \end{array} \right],  \ \ 
b^{\sigma}_{-} = \frac{e^{-i\phi/2}}{\sqrt{4\pi R}} \left[ \begin{array}{cc}
\cos(\theta/2) \\
-i \sin(\theta/2)   \end{array} \right].\]

Integrating over the sphere, the non-vanishing pairs between the above modes are given by 
\begin{equation}\label{pair}
\begin{split}
\int_{S^2} d^2x \sqrt{g} a^{\sigma}_{i}b^{\tau}_{j}B_{\sigma\tau} = \frac{R}{4} B_{ij},  \\ 
\int_{S^2} d^2x \sqrt{g} a^{\sigma}_{i}a^{\tau}_{j}\epsilon_{\sigma\tau} = \int_{S^2} d^2x \sqrt{g} b^{\sigma}_{i}b^{\tau}_{j}\epsilon_{\sigma\tau}  = -\frac{R}{4} \epsilon_{ij}.
\end{split}
\end{equation}

Restricted to the sector of these first order modes with $n=\pm 1$, we can expand the Grassmann fields defined in \eqref{fexp} as
\begin{eqnarray}\label{2dexp}
\xi_{a}^{\sigma \hat{\sigma}} = \xi_{a}^{i \hat{\imath}} \delta^{\hat{\sigma}}_{\hat{\imath}} (a_{i}^{\sigma} + i b_{i}^{\sigma})  + \tilde{\xi}_{a}^{i \hat{\imath}} \delta^{\hat{\sigma}}_{\hat{\imath}} (a_{i}^{\sigma} - i b_{i}^{\sigma}),  \\ \notag 
\lambda^{\sigma \hat{\sigma}} = \lambda^{i \hat{\imath}} \delta^{\hat{\sigma}}_{\hat{\imath}} (a_{i}^{\sigma} + i b_{i}^{\sigma})  + \tilde{\lambda}^{i \hat{\imath}} \delta^{\hat{\sigma}}_{\hat{\imath}} (a_{i}^{\sigma} - i b_{i}^{\sigma}). 
\end{eqnarray}
Here $\xi_{a}^{i \hat{\imath}}$,  $\tilde{\xi}_{a}^{i \hat{\imath}}$, $\lambda^{i \hat{\imath}} $ and $\tilde{\lambda}^{i \hat{\imath}}$ are Grassmann fields depending only on the coordinates of $M_3 = \Sigma\times I$. 
And we will shortly see that 
four independent fields among them will turn out to be massless in the three-dimensional action after reduction. 

With the fields expanded according to the above recipe, we can now do the dimensional reduction on $S^2$ for the fermion action. First, note that the Lagrangian can be written in three parts: 
\begin{equation}\label{3parts}
\rho_{m\Lambda} i (\slashed{\nabla}_{S^2})^{\Lambda}_{\Pi} \rho^{m\Pi}
+ \rho_{m\Lambda} i (\slashed{D}_{M_3})^{\Lambda}_{ \Pi} \rho^{m\Pi}
+ \rho_{m\Lambda} \frac{i}{2} (S-\slashed{T})^{\Lambda mn}_{\Pi } \rho^{\Pi}_{n}.
\end{equation}
After integrating over $S^2$ the first and the last terms become mass terms in the three-dimensional Lagrangian on $M_3$, while the second term turns into the kinetic part. 

To evaluate the action we decompose the spinor $\rho^{m\Lambda}$ as a linear combination (coefficients being Grassmann fields on $M_3$) of Killing spinors $a$, $b$ on $S^2$ using definitions \eqref{fexp} and \eqref{2dexp}. Further, we use the fact that the spinors $a$, $b$ are eigenfunctions of $\kappa\slashed{\nabla}_{S^2}$ as shown in \eqref{2deigen}. Finally, to integrate over $S^2$, we apply the scalar product formulae \eqref{pair} for the spinors $a$, $b$. For the first and last term in \eqref{3parts} this leads to an action
\begin{equation}
\begin{split}
\frac{i}{32\pi^2\tilde{R}}\int_{M_3}d^3x\sqrt{g_{M_3}}\ \text{Tr}& [B_{\hat{\imath}\hat{\jmath}}\epsilon_{ij}(\lambda^{i\hat{\imath}}\lambda^{j\hat{\jmath}} + \tilde\lambda^{i\hat{\imath}}\tilde\lambda^{j\hat{\jmath}} ) -B_{\hat{\imath}\hat{\jmath}}\epsilon_{ij}( \xi^{ai\hat{\imath}}\xi_{a}^{j\hat{\jmath}} + \tilde{\xi}^{ai\hat{\imath}}\tilde{\xi}_{a}^{j\hat{\jmath}}) \\
 +&\epsilon_{\hat{\imath}\hat{\jmath}}B_{ij}(\lambda^{i\hat{\imath}}\lambda^{j\hat{\jmath}} - \tilde\lambda^{i\hat{\imath}}\tilde\lambda^{j\hat{\jmath}} )], 
\end{split}
\end{equation}
while the second term becomes a kinetic term
\begin{equation}
\begin{split}
\frac{iR}{32\pi^2\tilde{R}}\int_{M_3}d^3x\sqrt{g_{M_3}}\ \text{Tr}& [ - 2B_{\hat{\imath}\hat{\jmath}} \epsilon_{ij}  \tilde{\xi}^{a i \hat{\imath}} D_a \lambda^{j\hat{\jmath}} -  2B_{\hat{\imath}\hat{\jmath}} \epsilon_{ij} \xi^{a i \hat{\imath}} D_a \tilde\lambda^{j\hat{\jmath}} ]. 
\end{split}
\end{equation}
Combining the above results, we finally reduce \eqref{3parts} to be
\begin{equation}\label{3dfaction}
\begin{split}
S_{f} = \frac{iR}{32\pi^2\tilde R} \int_{M_3}d^3x\sqrt{g_{M_3}}\ \text{Tr}[&  - 2B_{\hat{\imath}\hat{\jmath}} \epsilon_{ij}  \tilde\xi^{a i \hat{\imath}} D_a \lambda^{j\hat{\jmath}} \\ 
- &2 B_{\hat{\imath}\hat{\jmath}} \epsilon_{ij}  \xi^{a i \hat{\imath}} D_a \tilde\lambda^{j\hat{\jmath}} \\              -&\frac{1}{R}B_{\hat{\imath}\hat{\jmath}} \epsilon_{ij}(\xi^{a i \hat{\imath}}\xi_a^{j \hat{\jmath}} + \tilde\xi^{a i \hat{\imath}} \tilde\xi_a^{j \hat{\jmath}}) \\              +&\frac{4}{R}(B_{\hat{1}\hat{2}} \epsilon_{12} \lambda^{ 1 \hat{1}} \lambda^{ 2 \hat{2}}+B_{\hat{2}\hat{1}}\epsilon_{12}  \tilde\lambda^{ 1 \hat{2}} \tilde\lambda^{ 2 \hat{1}}) ].
\end{split}
\end{equation}
We can see that there are four massless field components which are
$\lambda^{1\hat{2}}$, $\lambda^{2\hat{1}}$, $\tilde\lambda^{1\hat{1}}$ and $\tilde\lambda^{2\hat{2}}$. 

Therefore, we find that in the lowest energy sector with eigenvalues equal to $\pm 1/R$, there are massless three-dimensional fields contained. But for the higher modes, by \eqref{eigen}, the massive terms given by the $\slashed{\nabla}_{S^2}$ term take a coefficient $n$ with $|n|>1$, whilst the last term of \eqref{3parts} still takes the same coefficient of the lowest energy sector, so there is no cancellation between the first and the last terms of \eqref{3parts}. Then higher modes with eigenvalues $> 1/R$ are massive of order $1/R$ and thus decoupled from the low-energy effective action when we take $R \to 0$. This leads us to concluding that the low-energy effective three-dimensional action is given by \eqref{3dfaction}. 

Here, we should point out that though the fermions $\lambda^{1\hat{2}}$, $\lambda^{2\hat{1}}$, $\tilde\lambda^{1\hat{1}}$ and $\tilde\lambda^{2\hat{2}}$  are massless, they couple to the massive fermions $\xi$ and $\tilde\xi$ in the kinetic terms; moreover, in the following we will see that massless and massive components of $\lambda$ and $\tilde\lambda$ couple to each other through Yukawa coupling. Consequently, we have to keep all the massive fermions in the action instead of just simply dropping them. 

\subsection{Interaction terms}
Now, to complete our reduction to the three-dimensional action, we continue to discuss the non-abelian interaction terms in the action. In our 5d Super-Yang-Mills theory, they take the form
\begin{equation}
S_{int} = \frac{1}{32\pi^2\tilde{R}} \int d^5x \sqrt{g}\ \text{Tr} (\rho_{m\Lambda}[\phi^{mn}, \rho^{\Lambda}_{n}] - \frac{1}{4}[\phi_{mn}, \phi^{nr}][\phi_{rs}, \phi^{sm}] - \frac{2}{3} S_{mn} \phi^{mr}[\phi^{ns}, \phi_{rs}]). 
\end{equation}
After integrating over the sphere, keeping only those terms that directly couple to the massless fermions, 
this term reduces to 
\begin{equation}
\begin{split}
S_{int}  =& \frac{R}{32\pi^2\tilde{R}} \int_{M_3} d^3x\sqrt{g_{M_3}}\  \text{Tr}[+ 2i\kappa^{z}_{\hat{1}\hat{1}}B_{12} \lambda^{1\hat{1}}[\phi_{z}, \lambda^{2\hat{1}}] 
+2i\kappa^{z}_{\hat{2}\hat{2}}B_{21} \lambda^{2\hat{2}}[\phi_{z}, \lambda^{1\hat{2}}] \\
&\ \ \ \ \ \ \ \ \ \ \ \ \ \ \ \ \ \ \ \ \ \ \ \ \ \ \ \ \ - 2i\kappa^{z}_{\hat{2}\hat{2}}B_{12} \tilde\lambda^{1\hat{2}}[\phi_{z}, \tilde\lambda^{2\hat{2}}]
- 2i\kappa^{z}_{\hat{1}\hat{1}}B_{21} \tilde\lambda^{2\hat{1}}[\phi_{z}, \tilde\lambda^{1\hat{1}}]\\
&\ \ \ \ \ \ \ \ \ \ \ \ \ \ \ \ \ \ \ \ \ \ \ \ \ \ \ \ \ +2i \epsilon_{\hat{1}\hat{2}}B_{21} \xi^{a2\hat{1}}[\phi_a, \lambda^{1\hat{2}}] 
+2i \epsilon_{\hat{2}\hat{1}}B_{12} \xi^{a1\hat{2}}[\phi_a, \lambda^{2\hat{1}}]   
\\
&\ \ \ \ \ \ \ \ \ \ \ \ \ \ \ \ \ \ \ \ \ \ \ \ \ \ \ \ \ -2i \epsilon_{\hat{2}\hat{1}}B_{11}  \tilde\xi^{a1\hat{2}} [\phi_a, \tilde\lambda^{1\hat{1}}] 
-2i \epsilon_{\hat{1}\hat{2}}B_{22}  \tilde\xi^{a2\hat{1}} [\phi_a, \tilde\lambda^{2\hat{2}}]  \\
&\ \ \ \ \ \ \ \ \ \ \ \ \ \ \ \ \ \ \ \ \ \ \ \ \ \ \ \ \  + 8\pi R[\phi_a, \phi_b][\phi_a,\phi_b] +  8\pi R[\phi_z, \phi_w][\phi_z,\phi_w]+ 16\pi R[\phi_z, \phi_a][\phi_z, \phi_a] \\
&\ \ \ \ \ \ \ \ \ \ \ \ \ \ \ \ \ \ \ \ \ \ \ \ \ \ \ \ \  + i\frac{16\pi}{3}\epsilon^{abc}\phi_a[\phi_b, \phi_c]  ].
\end{split}
\end{equation}

\subsection{Complex Chern-Simons theory as 3d effective theory}\label{S:cCS}
With the three-dimensional actions we obtained in Sections 4.1-4.4 for the massless modes by reduction, we are now ready to derive the 3d effective theory, which is a complex Chern-Simons theory, as we shall shortly see. 

To get the final form of the fermion action, the next step is to integrate out the massive modes (that directly couple to the massless modes). First, note that the fermion mass terms are proportional to $1/R$ in the action \eqref{3dfaction}, therefore their quantum fluctuations are suppressed in the path integral when we take the dimensional reduction limit $R\to 0$. Thus integrating out the massive modes can be done by replacing the massive fermions with solutions of their equations of motion:
\begin{equation}
\begin{split}
&\xi^{ii}_{a} = -R D_a \tilde\lambda^{ii},
\
 \xi^{i\slashed{i}}_{a} = - R[\phi_a, \lambda^{i\slashed{i}}] , \\
&\tilde\xi^{ii}_{a}=  -R[\phi_a, \tilde{\lambda}^{ii}],  \ 
\tilde\xi^{i\slashed{i}}_{a} = -R D_a \lambda^{i\slashed{i}}, \\
& \lambda^{11} = -\frac{R}{2}[\phi_z, \lambda^{12}]\kappa^{z}_{22}\epsilon^{21}, \ \lambda^{22} =- \frac{R}{2}[\phi_z, \lambda^{21}]\kappa^{z}_{11}\epsilon^{12}, \\
& \tilde\lambda^{12} = -\frac{R}{2}[\phi_z, \tilde\lambda^{11}]\kappa^{z}_{11}\epsilon^{12},   \
\tilde\lambda^{21} =  -\frac{R}{2}[\phi_z, \tilde\lambda^{22}]\kappa^{z}_{22}\epsilon^{21}, 
\end{split}
\end{equation}
where for making the notation more transparent, we use $ii$ to denote indexes 11 and 22,  and $i \slashed{i}$ to denote 12 and 21 in the equation (i.e. $i\neq \slashed{i}$). And note that in obtaining the above solutions, we have already decoupled the terms that do not directly couple to the massless fermions.  Plugging this solution into the fermion action, we get 
\begin{equation*}
\begin{split}
S_{f} = \frac{iR^2}{32\pi^2\tilde R} \int_{M_3}d^3x\sqrt{g_{M_3}}\ \text{Tr} [& 2 B_{12}\epsilon_{21} D_a\lambda^{21} D^a\lambda^{12} + 2 B_{12}\epsilon_{21} D_a \tilde\lambda^{22} D^a \tilde\lambda^{11}  \\
+& 2 B_{21}\epsilon_{12}[\phi_a,\lambda^{12}][\phi^a,\lambda^{21}] + 2 B_{21}\epsilon_{12}[\phi_a,\tilde\lambda^{11}][\phi^a,\tilde\lambda^{22}]  \\
-& B_{12}\kappa^z_{11} \kappa^w_{22}\epsilon^{21}[\phi_w,\lambda^{12}][\phi_z,\lambda^{21}] -B_{12}\kappa^z_{11} \kappa^w_{22}\epsilon^{21}[\phi_z,\tilde\lambda^{11}][\phi_w,\tilde\lambda^{22}] ].
\end{split}
\end{equation*}
By defining 
$\{\tilde\lambda^{11}, \lambda^{12}, \lambda^{21}, \tilde\lambda^{22} \} = \psi^{i\hat{\imath}}$, the above action can be rewritten as
\begin{equation}
\begin{split}
S_{f}= \frac{iR^2}{32\pi^2\tilde R} \int_{M_3}d^3x\sqrt{g_{M_3}}\ \text{Tr} [&   \epsilon_{ij} B_{\hat{\imath}\hat{\jmath}} D_a\psi^{i\hat{\imath}} D^a\psi^{j\hat{\jmath}} + \epsilon_{ij} B_{\hat{\imath}\hat{\jmath}}[\phi_a,\psi^{i\hat{\imath}}][\phi^a,\psi^{j\hat{\jmath}}]  \\
 -&\frac{1}{2}
(-\delta^{zw}\epsilon_{ij}B_{\hat{\imath}\hat{\jmath}}+i\epsilon^{zw}\epsilon_{ij}\epsilon_{\hat{\imath}\hat{\jmath}})[\phi_z,\psi^{i\hat{\imath}}][\phi_w,\psi^{j\hat{\jmath}}]], 
\end{split}
\end{equation}
while the action for bosons is 
\begin{equation}
\begin{split}
S_{b} = \frac{R^2}{4\pi \tilde{R}}  \int_{M_3} d^3x \sqrt{g_{M_3}}\ \text{Tr} 
[&F_{ab}F^{ab}   + 2\mD_a \phi^z \mD^a \phi_z + 2\mD_a \phi^b \mD^a \phi_b
 - i\frac{2}{R}\epsilon^{abc} \phi_a F_{bc} \\
    + & [\phi_a, \phi_b][\phi^a,\phi^b] +2[\phi_a, \phi_z][\phi^a,\phi^z] +  [\phi_z, \phi_w][\phi^z,\phi^w] \\
+ & i\frac{2}{3R}\epsilon^{abc}\phi_a[\phi_b, \phi_c] ].
\end{split}
\end{equation}
The above action is invariant under the following supersymmetric transformation:
\begin{equation}\label{susylaw}
\begin{split}
&\delta A_a = \zeta^{i\hat{\imath}} \epsilon_{ij}B_{\hat{\imath}\hat{\jmath}}D_a \psi^{j\hat{\jmath}}, \\
&\delta \phi_a = \zeta^{i\hat{\imath}} \epsilon_{ij}B_{\hat{\imath}\hat{\jmath}}[\phi_a, \psi^{j\hat{\jmath}}], \\
&\delta \phi_{w} =  \zeta^{i\hat{\imath}}[\phi_{z}, \psi^{j\hat{\jmath}}](-\delta^{zw}\epsilon_{ij}B_{\hat{\imath}\hat{\jmath}}+i\epsilon^{wz}\epsilon_{ij}\epsilon_{\hat{\imath}\hat{\jmath}}), \\
&\delta\psi^{i\hat{\imath}} =16\pi i\epsilon^{wz}\kappa^{\hat{\imath}}_{\hat{\jmath}}B^{i\hat{\jmath}}[\phi_z, \phi_w],
\end{split}
\end{equation}
where $\zeta^{i\hat{\imath}}$ is the Grassmann coefficient of supersymmetry transformation.
   
At the end of this section, we will see that the action defined above can eventually give rise to the (imaginary part of the) complex Chern-Simons theory. 

\subsection*{Q-exact terms}
As elaborated in the introduction, we focus on the supersymmetric invariant sector of the supersymmetric theory, whereby we consider only the $Q$-invariant observables. Under this restriction, the $Q$-exact term in the action can be freely subtracted without affecting the physics, i.e. without affecting the expectation value of the $Q$-invariant observables \cite{witten1988}.        

Let us define 
\begin{equation} 
\Xi_{i\hat{\imath}} = \text{Tr}(\psi_{i\hat{\jmath}}\kappa^{\hat{\jmath}}_{\hat{\imath}}[\phi_z, \phi_w]\epsilon^{zw}).
\end{equation}
Then, by \eqref{susylaw}, we have
\begin{equation} \label{qexact}
\delta^{i\hat{\imath}} \Xi_{i\hat{\imath}} =    (-\delta^{zw}\epsilon_{ij}B_{\hat{\imath}\hat{\jmath}}+i\epsilon^{zw}\epsilon_{ij}\epsilon_{\hat{\imath}\hat{\jmath}}) [\phi_z, \psi^{i\hat{\imath}}][\phi_w, \psi^{j\hat{\jmath}}] + 16\pi i [\phi_z, \phi_w][\phi^z,\phi^w].
\end{equation}
Here we find that the above $Q$-exact terms are just the non-linear terms for fields $\phi_z$ and $\psi$ in our action. And by the $Q$-exact property, we subtract them from the action freely for simplifying our following discussion.

\subsection*{Ghost terms}
As our purpose is to obtain a 3d theory depending only on $A_a$ and $\phi_a$, we need to integrate out the terms containing the fields $\psi^{i\hat{\imath}}$ and $\phi_z$ in the action. After subtracting the $Q$-exact terms \eqref{qexact} in the action, these terms -- which we denote as $S_{gf}$ -- are left to be
\begin{equation}
\begin{split}
S_{gf} = \frac{R^2}{32\pi^2\tilde R} \int_{M_3} d^3x\sqrt{g_{M_3}}\ \text{Tr}[&16\pi\mD_a \phi^z \mD^a \phi_z  + B_{\hat{\imath}\hat{\jmath}}\epsilon_{ij} D_a\psi^{i\hat{\imath}} D^a\psi^{j\hat{\jmath}}  \\
+ &16\pi[\phi_a, \phi_z][\phi^a,\phi^z] + B_{\hat{\imath}\hat{\jmath}}\epsilon_{ij}[\phi_a,\psi^{i\hat{\imath}}][\phi_a,\psi^{j\hat{\jmath}}] ].
\end{split}
\end{equation}
In the following we can see that the above fermions can be reinterpreted as Faddeev-Popov ghosts of gauge fixing an emergent non-compact gauge symmetry.

As the above action is quadratic, by performing the path integral, up to a normalization factor, we get 
\begin{equation}\label{determ}
\int \mathcal{D} \phi_z \mathcal{D} \psi e^{-S_{gf}} = \text{det}(D_a^{2} + (\text{ad}_{\phi_a})^2), 
\end{equation}
where $\text{ad}_{\phi_a}$ denotes the operator of $\phi_a$ in the adjoint representation acting on fields. And note that to get this result, we used the fact that the functional integral over two real boson fields $\phi_z$ just cancels out the functional integral over two of the fermions fields $\psi$, since the action forms for the bosons and fermions are identical. 

Next, we shall see that the above path integral determinant \eqref{determ} -- together with one gauge fixing term in the action --  manifestly gauge fixes the non-compact part of the complex gauge symmetry which is defined in the following. 

\subsection*{Complex Chern-Simons Theory}
Up to this point, we can summarize our result as 
\begin{equation}\label{partition}
\begin{split}
\int \mathcal{D} A_a\mathcal{D}\phi_a  \text{det}(D_a^{2} + (\text{ad}_{\phi_a})^2) e^{-S(\phi_a, A_a)},
\end{split}
\end{equation}
with
\begin{equation}
\begin{split}
S(\phi_a, A_a) = \frac{R^2}{4\pi \tilde{R}}  \int_{M_3} d^3x \sqrt{g_{M_3}}\ \text{Tr} 
[&F_{ab}F^{ab}   + 2\mD_a \phi_b \mD^a \phi^b
 - i\frac{2}{R}\epsilon^{abc} \phi_a F_{bc} \\
 + &  [\phi_a, \phi_b][\phi^a,\phi^b]  \\
+ & i\frac{2}{3R}\epsilon^{abc}\phi_a[\phi_b, \phi_c] ]. 
\end{split}
\end{equation}

Then, 
by defining the complex gauge field as 
\begin{equation}\label{eq:defcA}
\mA_a = A_a + i \phi_a,
\end{equation}
the above action can be rewritten as 
\begin{equation}\label{csaction}
\begin{split}
S(\phi_a, A_a) &= S_{CS} + \frac{R^2}{2\pi \tilde{R}} \int d^3x \mathcal{F}_{ab}\bar{\mathcal{F}}^{ab}  - \frac{R^2}{2\pi \tilde{R}} \int d^3x (D_a \phi^a)^2,\ \ \text{with} \\
S_{CS} &= -\frac{R}{4\pi\tilde R} \int \text{Tr} (\mA \wedge d\mA + \frac{2}{3} \mA \wedge \mA \wedge \mA) + \frac{R}{4\pi\tilde R}\int \text{Tr} (\bar \mA \wedge d \bar \mA + \frac{2}{3}\bar \mA \wedge \bar \mA \wedge \bar A). 
\end{split}
\end{equation}

Since under the complexified gauge transformation $\bold{g}_{\mathbb{C}}$, the divergence term $(D_a \phi^a)^2$ is not invariant, thus it can manifest itself as a gauge fixing term. 
Under the non-compact part of the $\bold{g}_{\mathbb{C}}$ gauge transformations, the fields $\phi$ and $A$ transform as
\begin{equation}
\delta_{\bold{g}} \phi_a = D_a g,\quad \delta_{\bold{g}} A_a = [g, \phi_a]. 
\end{equation}
By these transformations, we have 
\begin{equation}
\delta_{\bold{g}}(D_a \phi^a) = D^2 g + (\text{ad}_\phi)^2 g. 
\end{equation}
The above equation indicates that the Faddeev--Popov determinant for this gauge fixing term is exactly the determinant in the modified measure \eqref{partition}. Therefore the non-compact part of the gauge group is nicely gauge fixed. Furthermore, in the limit $R\to 0$, the second Yang--Mills term in \eqref{csaction} is negligible and can be simply dropped. We then recognize our action to be the imaginary part of the complex Chern-Simons theory, where the real part of the level $k=0$ and the imaginary part $s=i\frac{R}{\tilde{R}}$.

\section{Reduction to 2d theory and 4d-2d duality}\label{S:Toda}
\subsection{Complex Toda}
Since the complex Chern-Simons theory we got is on the manifold $M_3 = \Sigma\times I$, we can further reduce the complex Chern-Simons to a two-dimensional theory of edges modes on $\Sigma$.  

Here we follow the discussion in \cite{Elitzur:1989nr}. On the boundary, (denoting $p$ and $\bar{p}$ as the complex coordinates on $\Sigma$, while $u$ the coordinate on $I$,) the gauge field $\mathcal{A}_{\bar{p}}$ can serve as a Lagrange multiplier, and integrating it out in the path integral leads to the constraint $\delta{(\mathcal{F}_{pu})}$, i.e. $\mathcal{F}_{pu}$ being flat. Under this condition, the gauge fields $\mathcal{A}=\left(\mathcal{A}_p,\mathcal{A}_u\right)$ take the form: 
\begin{equation}
\mathcal{A}=G^{-1}\tilde{d} G,
\end{equation}
where $G$ depends on $p,\bar{p},u$ and $\tilde{d}$ is the exterior derivative on $(p,u)$. Similarly, for their complex conjugate fields, we have $\bar{\mathcal{A}}=\bar{G}^{-1}\tilde{d} \bar{G}$. Then in terms of the group fields $G$ and $\bar{G}$, the complex Chern-Simons theory reduces to the WZW model 
\begin{equation}\label{WZW}
S=\frac{k+is}{2}I_{WZW}(G)+\frac{k-is}{2}I_{WZW}(\bar{G}),
\end{equation}
with \begin{equation*}
I_{WZW}(G)=\frac{1}{2\pi}\int_{\Sigma}\Tr(G^{-1}\partial_p G G^{-1}\partial_{\bar{p}} G)+ \frac{1}{6\pi}\int_{\Sigma\times I}\Tr(G^{-1}dG\wedge G^{-1}dG\wedge G^{-1}dG).
\end{equation*}
Here, the coupling constant ($k, is$) is the same as the coupling constant of the complex Chern-Simons theory. So in the above WZW action, $k=0$ and $s= i\frac{R}{\widetilde{R}}$.

Next, we will see that on the boundary, besides the flatness constraint $\mathcal{F}_{pu}=0$, the gauge fields are actually further constrained by additional boundary conditions. 

Our Chern-Simons theory is deduced from the six-dimensional (2,0) theory which is the low-energy description of the world-volume theory of multiple M5-branes and hence the boundary conditions imposed on Chern-Simons theory should be also inherited from those imposed on the latter theory. Considering the dimensional reduction of the $S^1$ circle as the collapse of the eleventh-dimensional circle in M-theory, the M5-brane reduces to the D4 brane 
\cite{lambert_m5-branes_2011, douglas_d5_2011}. This reduction is consistent with our Chern-Simons theory where the complex gauge field $\mathcal{A}=A+i\phi$ is composed of the world-volume gauge field and a triplet of twisted scalars of the D4 brane theory. Hence,  the boundary conditions imposed on M5-brane are analogous to  those on D4-brane  ending on D6-brane. Since the Nahm pole bondary condition -- which has been studied in detail in \cite{gaiotto_supersymmetric_2009, gaiotto_s-duality_2009} in the context of the $D3-D5$ system -- is a general property of the $Dp-D(p + 2)$ system \cite{witten_fivebranes_2011} for any $p$, we should further impose the Nahm pole boundary conditions on our complex Chern-Simons theory. Therefore,
the flat connections should be further restricted to \cite{Cordova:2016cmu}
\begin{equation}
\mathcal{A}=\frac{du}{u}H +\frac{dp}{u}T_+ + \sum^{n-1}_{j=1} \chi_j(u)(T_-)^ju^j dp,
\end{equation}
where $H$ is the Cartan generator and $T_+$, $T_-$ are the raising and lowering operators as introduced in \cite{Cordova:2016cmu}. Given the Nahm pole boundary condition, the WZW model can subsequently reduce to the complex Toda theory \cite{Cordova:2016cmu}:
\begin{equation}\label{complex_liouville}
S=\frac{k+is}{4\pi}\int ( C_{ij}\partial \Phi^i\bar{\partial}\Phi^j +\sum_i \exp (C_{ij}\Phi_j))  +\frac{k-is}{4\pi}\int (C_{ij}\partial\bar{\Phi}^i\bar{\partial}\bar{\Phi}^j + \sum_i\exp(C_{ij}{\bar{\Phi}_j})),
\end{equation}
where $\Phi$ is related to the WZW group field $G$ by an exponential change of variables and contains $r$ components where $r$ is the rank of Lie group algebra $\mathfrak{g}$; 
$C_{ij}$ is the Cartan matrix of $\mathfrak{g}$. The complex Toda theory is equivalent to a complex WZW model with an additional constraint on the currents imposed by the Nahm pole boundary conditions. Note that in our case $k=0$. 


The major property of both our complex 3d and 2d theories is that the real part of the complex coupling constant, $k$, is zero. Following the discussion of \cite{Dimofte:2011py} by Dimofte et al., we first note that the Chern-Simons path integral on M is related to a wavefunction in a boundary Hilbert space $\mathcal{H}_{\partial M}$. And $\mathcal{H}_{\partial M}$ is given by the quantization of the classical phase space associated to the boundary: 
\begin{equation}
\mathcal{P}_{\partial M} \simeq \{\text{flat connections}\  \mathcal{A} = \frac{d\sigma}{\sigma}H + \frac{du}{\sigma} T_{+} + \sum^{n-1}_{j=1} \chi_{j}(u)(T_{-})^j\sigma^{j}du \}.  
\end{equation}
In the above equation the Nahm pole boundary condition is imposed. Notably, this quantization depends critically on the relative values of $k$ and $s$, which determine the real symplectic structure of the phase space. The quantization for the $k=0$ case is elaborated in \cite{Dimofte:2011py}.  

Next, to discuss the two dimensional complex Toda theory with $k=0$, we first note that for the $k\geq 1$ cases, the previous work by Cordova and Jafferis \cite{Cordova:2016cmu} showed that the complex Toda theory is dual to the para-Toda theory plus a decoupled coset model. In \cite{Cordova:2016cmu} the authors considered the six-dimensional (2,0)  theory on $S^4_l/{\mathds{Z}_k}\times \Sigma$ ($k\geq 1$), then they showed that its reduction to two-dimensional $\Sigma$ gives rise to a complex Toda theory. 
However, in their derivation from the complex Toda to the para-Toda theory, $k$ can never be zero; and the useful hint guiding them to the final result is the known generalized AGT correspondence stating that the $\mathfrak{su}(n)$ (2,0) theory on $S^4_l/{\mathds{Z}_k}$ can be described in terms of a para-Toda theory plus a decoupled coset model \cite{nishioka_para-liouville/toda_2011}, where $S^4_l/\mathds{Z}_k$ also can
not be generalized to include the $k=0$ case. Therefore, we can not do a simple analogue in our case, thus for now the relation between the complex Toda theory of $k=0$ (i.e. the imaginary part of the complex Toda theory) and the real Toda theory is not yet clear to us. 

\subsection{4d-2d duality}
 In our discussion hitherto, we have shown that by the reductions on $S^1$ and $S^2$, the 6d (2,0) SCFT on $S^2\times\Sigma\times I \times S^1 $ results in a complex Chern-Simons theory on $\Sigma \times I$, and then by the reduction on $I$ under the Nahm pole boundary condition, this 3d complex Chern-Simons gives rise to a complex Toda theory of $k=0$ on $\Sigma$. 

Here, we notice the following three facts -- (I) the complex Toda CFT is conformal on $\Sigma$, so we can freely resize $\Sigma$ without altering the theory; (II) the complex Toda CFT is obtained by reducing the 6d (2,0) SCFT, in a certain supersymmetric low energy sector, down to $\Sigma$; and (III) we topologically twist along $\Sigma$. Because of (I) and (II), the aforementioned supersymmetric low energy sector is insensitive to rescalings of $\Sigma$. Hence, with respect to this supersymmetric low energy sector, our reduction of the 6d theory down to $\Sigma$ -- which can be regarded as scaling $\Sigma$ up to be much larger than $S^2 \times I \times S^1$ -- would be equivalent to a reduction of the 6d theory down to $S^2 \times I \times S^1$  -- which can be regarded as scaling $\Sigma$ down to be much smaller than $S^2 \times I \times S^1$. Furthermore, because of (III), we would not break any of the eight (conformally-extended) supercharges when we scale $\Sigma$ down -- in other words, the reduction of the 6d theory down to $S^2 \times I \times S^1$ should result in a 4d $\mathcal{N}=2$ supersymmetric theory. Altogether, this means that with respect to this supersymmetric low energy sector, our results imply a 4d-2d duality between four-dimensional $\mathcal{N}=2$ supersymmetric theory with boundary on $S^2 \times I \times S^1$ (where the $\mathcal{N}=2$ supersymmetries are broken to $\mathcal{N}=1$ at the boundary by the Nahm pole boundary condition\footnote{Note that since the Nahm pole boundary condition is a general property of $Dp-D(p+2)$ system~\cite{witten_fivebranes_2011}, for our discussion this boundary condition should also be imposed on the 4d theory. (The Nahm pole boundary condition for the $N=4$ super-Yang-Mills has been studied in detail in~\cite{gaiotto_supersymmetric_2009,gaiotto_s-duality_2009} in the context of $D3-D5$ system.)}) and two-dimensional complex Toda theory on $\Sigma$.

\section{Acknowledgements}
We would like to thank Junya Yagi and Masaya Yata for helpful discussions. The work is supported by NUS Tier 1 FRC Grant R-144-000-316-112. P.V. would also like to acknowledge the ERC Starting Grant no. 335739 ``Quantum Fields and Knot Homologies" funded by the European Research Council under the European Union's Seventh Framework Programme and the Foundation for Polish Science by which he is supported since 1 October 2016.

\appendix

\section{Conventions}\label{conventions}
Our index conventions are shown as follows:
\begin{center}
\begin{tabular}{rcccc}
\hline
Lorentz indices & 6d & 5d & $S^2$ &$\Sigma\ \times\ I$\\ 
\hline
Vector (curved) & $\underline{\mu }, \underline{\nu}$ & $\mu, \nu$ & $\theta, \phi$  & $x,y\ \ \ u$\\ 
Vector (frame) & $\underline{A}, \underline{B}$ & $A, B$ & $z,w$ & $(a,b)$\\
Spinor & -- & $\Lambda,\Pi$ & $\sigma,\tau$  & $\alpha, \beta \ \ \ -$\\
\end{tabular}
\end{center}

\begin{center}
\begin{tabular}{ccc}
\hline 
R-symmetry group & $SO(5)$ & $Sp(4)$ \\
\hline
R-symmetry indices & $\hat{A}, \hat{B}$ & $m,n$ \\
\end{tabular}
\end{center}
Following the above index conventions, the background fields in $SO(5)_R$ representation convert to $Sp(4)_R$ representation as follows:
\begin{align}
T^{mn}_{BC}= T_{\hat{A}BC}\left( \Gamma^{\hat{A}}\right)^{mn}, & \quad  V_A^{mn}= V_{A\hat{B}\hat{C}}\left( \Gamma^{\hat{B}\hat{C}}\right)^{mn}, \\
 S^{mn}= S_{\hat{B}\hat{C}}\left( \Gamma^{\hat{B}\hat{C}}\right)^{mn}, & \quad D^{mn, rs}= D_{\hat{A}\hat{B}}\left( \Gamma^{\hat{A}}\right)^{mn} \left( \Gamma^{\hat{B}}\right)^{rs}. 
\end{align}

\section{5d gamma matrices and operations on spinors}\label{App:GammaMat}
\subsection*{5d gamma matrices} 

Here we list explicit formulae that we use for gamma matrices in five dimensions with Euclidean signature
\begin{align}\label{eq:GammaMatrices}
\Gamma^1=\mathds{1}\otimes\kappa^1, && \Gamma^2=\mathds{1}\otimes\kappa^2, && \Gamma^3=\kappa^1\otimes\kappa, && \Gamma^4=\kappa^2\otimes\kappa, && \Gamma^5=\kappa\otimes\kappa,
\end{align}
where $\{\kappa^1,\kappa^2,\kappa\}$ are the standard Pauli matrices:
\begin{align}
\kappa^1=\left(
\begin{array}{cc}
0 & -i \\
i & 0
\end{array}\right),
&&
\kappa^2=\left(
\begin{array}{cc}
0 & 1 \\
1 & 0
\end{array}\right),
&&
\kappa=\left(
\begin{array}{cc}
1 & 0 \\
0 & -1
\end{array}\right).
\end{align}
They satisfy the Clifford algebra
\begin{equation} \label{gamma}
\left\{\Gamma_A,\Gamma_B\right\}=2g_{AB}\mathds{1}_{4\times 4}.
\end{equation}
The same set of gamma matrices is used for both the Lorentz group $SO(5)_L$ and the R-symmetry group $SO(5)_R$.

The symmetric $B$ matrices appearing in the text take the form
\begin{equation}
B_{\sigma\tau}=B^{\sigma\tau}=B_{\hat{\sigma}\hat{\tau}}=B^{\hat{\sigma}\hat{\tau}}:=\kappa^2=\left(
\begin{array}{cc}
0 & 1 \\
1 & 0
\end{array}\right).
\end{equation}

\subsection*{Raising and lowering spinor indices} 

All five dimensional spinors in the text are elements of the spinor representation $Spin(5)$, as well as of the defining $\mathbb{C}^4$ representation of $Sp(4)$. The latter is the R-symmetry representation inherited from the six dimensional $(2,0)$ theory (as it survives the dimensional reduction unbroken). To make the representation content manifest, we write the spinor as $\rho^{\Lambda m}$, where $\Lambda$ labels the elements of the spinor representation of $Spin(5)$ while $m$ labels those of the defining representation of $Sp(4)$. However, $Spin(5)\simeq Sp(4)$ (in particular the spinor representation of Spin(5) is isomorphic to the defining representation of $Sp(4)$), so we may treat both indices at equal footing and we adopt this convention throughout the paper. Having said that, let us present the following discussion just for one of the two equivalent indices, for instance $m$ of $Sp(4)$. The elements of $\mathbf{4}\;(\rho^m)$ are isomorphic to elements of the dual representation $\mathbf{4}^\vee\;(\rho_m)$ via the pairing
\begin{align}
\rho^m=\Omega^{mn}\rho_n,
\end{align}
where we choose $\Omega^{mn}$ in accordance with the R-symmetry breaking $Sp(4)\to Sp(2)\times Sp(2)$ as
\begin{align}
\Omega^{mn}=\Omega^{(\hat{\alpha}\hat{\sigma})(\hat{\beta}\hat{\tau})}=\epsilon^{\hat{\alpha}\hat{\beta}}\otimes B^{\hat{\sigma}\hat{\tau}}=\begin{pmatrix} 0 & 1 \\ -1 & 0 \end{pmatrix}\otimes \begin{pmatrix} 0 & 1 \\ 1 & 0 \end{pmatrix}.
\end{align}

\section{Five-dimensional SYM in supergravity background}\label{App:SUGRA}
The action for the vector multiplet in the supergravity background consists of four terms 
\begin{equation}
S= S_A + S_\varphi + S_\rho + S_{int},
\end{equation}
which are given as
\begin{align}
S_A & =\frac{1}{8\pi^2}\int \Tr \left( \alpha F\wedge\star F + C\wedge F \wedge F\right), \label{eq:SA} \\
S_\varphi &= \frac{1}{32\pi^2}\int d^5 x \sqrt{|g|}\alpha\Tr \left( \mathcal{D}_A\varphi^{mn}\mathcal{D}^A \varphi_{mn} - 4\varphi^{mn}F_{AB}T^{AB}_{mn} - \varphi^{mn}(M_\varphi)^{rs}_{mn}\varphi_{rs}\right), \\
S_\rho &= \frac{1}{32\pi^2}\int d^5 x \sqrt{|g|}\alpha\Tr\left(\rho_{m\Lambda}i\slashed{\mathcal{D}}^\Lambda_\Pi \rho^{m\Pi} + \rho_{m\Lambda}(M_\rho)^{mn\Lambda}_{\Pi}\rho^{\Pi}_n\right), \\
S_{int} & = \frac{1}{32\pi^2}\int d^5 x \sqrt{|g|}\alpha\Tr\left( \rho_{m\Lambda}[\varphi^{mn},\rho^{\Lambda}_n] - \frac{1}{4}[\varphi_{mn},\varphi^{nr}][\varphi_{rs},\varphi^{sm}] - \frac{2}{3}S_{mn}\varphi^{mr}[\varphi^{ns}, \varphi_{rs}]\right) \label{eq:Sint},
\end{align} 
where the covariant derivatives act on $\varphi$ and $\rho$ as
\begin{align}
\begin{aligned}
\mathcal{D}_\mu \varphi_{mn} & = \left(\partial_\mu - \partial_\mu \text{log}(\alpha)\right)\varphi_{mn} -V^r_{\mu[m}\varphi_{n]r} + [A_\mu, \varphi_{mn}], \\
\mathcal{D}_\mu \rho^m &= \left(\partial_\mu -\frac{3}{2}\partial_\mu \text{log}(\alpha) +\frac{1}{4}\omega^{BC}_\mu\Gamma_{BC} \right) \rho^m -\frac{1}{2}V^m_{\mu n}\rho^n + [A_\mu,\rho^m].
\end{aligned}
\end{align}
And the supergravity induced mass matrices are 
\begin{align}
\begin{aligned}
(M_\varphi)^{rs}_{mn} & = \left[\left(\frac{1}{20\alpha^2}G_{AB}G^{AB}-\frac{\mathcal{R}}{5}\right)\delta^r_m\delta^s_n + \frac{1}{2}\left( S^r_{[m}S^s_{n]} - S^s_t S^t_{[m}\delta^r_{n]}\right) -\frac{1}{15} D^{rs}_{mn} - T^{AB}_{mn} T^{rs}_{AB}\right], \\ 
(M_\rho)^{mn\Lambda}_{\Pi} & =\left[\frac{1}{2}S^{mn}\delta^\Lambda_\Pi + \frac{1}{8\alpha}G_{AB}(\Gamma^{AB})^\Lambda_\Pi\Omega^{mn} - \frac{1}{2}T^{mn}_{AB}(\Gamma^{AB})^\Lambda_\Pi\right].
\end{aligned}
\end{align}
Here, $\mathcal{R}$ is the Ricci scalar curvature of the five-dimensional metric. This holds if the dilaton is constant, which is true for our geometry. Then $\omega_\mu^{AB}$ is just the usual torsion-free spin connection of Riemannian geometry
\begin{equation}
\mathcal{R} = e^\mu_A e^\nu_B\left(2\partial_{[\mu}\omega^{AB}_{\nu]}+2\omega^{AC}_{[\mu}\omega_{\nu]C}^B\right).
\end{equation}

\bibliography{bib_pv} \bibliographystyle{JHEP}
\end{document}